\documentclass[a4paper,10pt]{article}
\usepackage{graphicx}
\usepackage{amssymb}
\usepackage{amsmath}
\usepackage{nicefrac}
\usepackage{dcolumn}
\usepackage{multirow}
\usepackage{cite}
\usepackage{mathrsfs}
\usepackage{bm}
\usepackage[usenames,dvipsnames]{xcolor}
\usepackage[colorlinks=true,citecolor=Blue,linkcolor=RubineRed,urlcolor=Blue]{hyperref}
\usepackage{bbm}
\usepackage{hyperref}
\usepackage{comment}
\usepackage{amsfonts}
\usepackage{float}
\usepackage{caption}
\usepackage{subcaption}

\usepackage{xcolor}
\usepackage{soul}
\setstcolor{red}

\begin{document}

\huge

\begin{center}
Average-atom Ziman resistivity calculations in expanded metallic plasmas: effect of mean ionization definition
\end{center}

\vspace{0.5cm}

\large

\begin{center}
Nadine Wetta$^{a,}$\footnote{nadine.wetta@cea.fr} and Jean-Christophe Pain$^{a,b}$
\end{center}

\normalsize

\begin{center}
\it $^a$CEA, DAM, DIF, F-91297 Arpajon, France\\
\it $^b$Universit\'e Paris-Saclay, CEA, Laboratoire Mati\`ere sous Conditions Extr\^emes,\\
\it 91680 Bruy\`eres-le-Ch\^atel, France
\end{center}

\vspace{0.5cm}

\begin{abstract}
We present calculations of electrical resistivity for expanded boron, aluminum, titanium and copper plasmas using the Ziman formulation in the framework of the average-atom model. Our results are compared to experimental data, as well as with other theoretical calculations, relying on the Ziman and Kubo-Greenwood formulations, and based on average-atom models or quantum-molecular-dynamics simulations. The impact of the definition of ionization, paying a particular attention to the consistency between the latter and the perfect free electron gas assumption made in the formalism, is discussed. We propose a definition of the mean ionization generalizing to expanded plasmas the idea initially put forward for dense plasmas, consisting in dropping the contribution of quasi-bound states from the ionization due to continuum ones. It is shown that our recommendation for the calculation of the quasi-bound density of states provides the best agreement with measurements. 
\end{abstract}

\section{Introduction}

Many fields of physics are related to the study of atomic properties of warm dense matter. More specifically, there is a constant need to improve models for accurate equations of state and transport coefficients. Among the latter, the electrical resistivity is of particular importance for the study of the intense energy-flux interactions with matter that occur in laser-fusion experiments \cite{glenzer2010,hurricane2014}. Many experimental methods have been developed to investigate transport properties. Among them, high intensity heavy ion beams are used to probe large volumes of high-energy matter \cite{udrea2006}, pulsed electrical currents to explode wires or to rapidly heat foils \cite{krisch1998,desilva1999}, highly bright x-ray beams \cite{sperling2015} or intense laser beams \cite{milchberg1988} to heat solid targets at constant volume, while the ``isochoric plasma closed vessel'' facility allows for the study of expanded materials under isochoric conditions \cite{renaudin2002,recoules2002}.

The direct current (dc) electrical resistivities are currently calculated within the Ziman-Evans (ZE) theory \cite{ziman1961,evans1973}, which describes the scattering of free electrons in a metal by an ion. Among the physical quantities needed, those necessary to build the scattering cross-section can be provided by quantum-molecular-dynamics (QMD) or average-atom (AA) methods. 

Standard average-atom codes, including our {\sc Paradisio} one \cite{penicaud2009,wetta2020} and the {\sc Purgatorio} code \cite{wilson2006}, are based on Liberman's {\sc Inferno} atom-in-jellium model of matter \cite{liberman1979}. The strong assumption of the {\sc Inferno} model is the muffin-tin approximation, consisting in considering that beyond the ionic cell, the electron density is constant and equal to the jellium density. Such an assumption, coming from solid-state theory, has important consequences on the model. The global neutrality of the system boils down to the neutrality of the ionic cell. The potential is zero outside the cell. In that sense, the atomic sphere is separated from the surrounding jellium, and all the calculations are performed in the ionic (Wigner-Seitz) cell. {\sc Inferno}-based average-atom methods are renowned for giving rather accurate results at reasonable computational cost. These results can be used in the Ziman-Evans formalism to calculate electrical resistivities.

The Ziman-Evans formula expresses the electrical resistivity in terms of electron scattering phase-shifts, the ion charge (also called the mean ionization) $Z^*$ and the ion-ion structure factor $S(q)$. Two issues arise then when results from average-atom codes are used. The ion-ion structure factor is not provided by {\sc Inferno}-type average-atom models, which are single-atom ones, and must be obtained separately. Many models are available for this quantity, each of them having its own range of validity. Interpolations between two limiting models, like in \cite{sperling2017}, are often necessary. A similar problem appears for $Z^*$, which is not unambiguously defined since it is not a quantum-mechanical observable, and for which many formula are in competition.

Recently, we published an article (referred to as ``Paper I'' in the following) \cite{wetta2020} on the calculation of electrical conductivity of aluminum at solid density $\rho=2.7$ g/cm$^3$, using the Ziman-Evans approach in the framework of the average-atom code {\sc Paradisio}. Our choice of aluminum at this density was dictated by the great number of experimental and theoretical results available for comparisons. We investigated a large temperature range, going from ambient one up to 100 eV, thus covering solid state, melting, liquid and plasma states. In this study, $Z^*$ was defined as the number of electrons in the continuum states, a choice that can be justified by the fact that is the one that recovers the $Z^*=3$ value ({\it i.e.} Al$^{+3}$ charge) in the solid and liquid states. Having thus solved the problem of the definition of $Z^*$, we focused our attention on the ion-ion structure factor, in order to, on one hand, to reproduce the resistivity jump at melting, and on the other hand, to propose a continuous model for $S(q)$ suitable from the liquid to the plasma states. Our first goal was achieved by extending to the liquid a concept inherited from solid state theory, and consisting in dropping elastic scattering contributions from the total ion-ion structure factor. This led us to derive a formula for the correction that must be applied to Ziman-Evans's resistivity above melting. The correction is the highest at melting and vanishes progressively as temperature increases. Our second aim, that was to find a unique description from liquid to plasma, using the same model for the total $S(q)$, is therefore also achieved.

In an other article (``Paper II'') \cite{wetta2022}, we presented some complementary results to the previous work. Still for solid-density aluminum, the impact of different definitions for $Z^*$ and different models for $S(q)$ on electrical resistivity was investigated. A direct effect of the definition retained for $Z^*$ on the Ziman-Evans resistivity through the factor ${1/Z^*}^2$ (ZE formula will be given in the main text) is obvious. We showed that $Z^*$ also impacts indirectly the results through the structure factor and the chemical potential, and that this counteracts the direct effect. Most probably by chance, the compensation is nearly complete in the case of dense aluminum. Convinced that compensation is not the rule, we believe in the necessity to, at least, frame the possible values of $Z^*$, and, ideally, find a global definition for it, in the spirit of the work we have done on the structure factor.

This is the main aim of the present paper. We calculate electrical resistivities for expanded boron, aluminum, titanium and copper, within the Ziman-Evans theory, in the framework of our average-atom code {\sc Paradisio}. The densities (in the order of a few times $\rho_\mathrm{solid}/10$) and temperatures (ranging from $10^4$ K to $4\ 10^4$ K) of our calculations are favorable to large differences in $Z^*$ according to its definition, and correspond to the conditions of the experiments performed on the EPI ``isochoric plasma close vessel'' facility \cite{clerouin2012}, to which we compare our results. We also confront the latter to published theoretical values \cite{clerouin2012,korobenko2005,krisch1998,desjarlais2002}.

Ziman's formalism imposes electrical neutrality of the system, {\it i.e.} $Z^*$ must correspond to the scattered electron gas charge. The latter electron gas is also assumed to be perfectly uniform and free. The definition of $Z^*$ should be consistent with this. We therefore, all along the paper, pay particular attention to the consistency between definition of the ion charge and the perfect free-electron-gas assumption underlining the Ziman-Evans formalism.

The Ziman-Evans formula, and the different ways to define the mean ionization from the code are described in section II. The above mentioned comparisons of our results to experiments and other theoretical works is presented in section III. A definition for the mean ionization $Z^*$ emerges then among the different ones we probed. It yields the best possible agreement with experiments and theories, given the approximations underlying the {\sc Inferno} model, and extends, on a natural way, to low densities an expression already proposed for normal or dense ones. In section \ref{sec4}, we try to go beyond the ideal free electron gas assumed in Ziman's formalism by the introduction of a real electronic density of states (DOS) in the formula, before ending, in section \ref{sec5}, on the conclusions that can be drawn from the present work. Finally, a description of {\sc Paradisio} is proposed in the Appendix.\\

\section{Calculation of electrical resistivity in the framework of the average-atom model}\label{sec2}

\subsection{The Ziman-Evans formulation}

The atomic units are used throughout this paper, {\it i.e.} $\hbar=m=e=1$ where $\hbar$, $m$, and $e$ denote respectively the Planck constant, the electron mass and its electrical charge. The Boltzmann constant $k_B$ is also set to unity.

Ziman's formulation \cite{ziman1961} describes, within the linear response theory, the acceleration of free electrons in a metal and their scattering by an ion:
\begin{equation}\label{eta}
\eta=-\dfrac{1}{3\pi {Z^*}^2 n_i} \int_0^\infty \dfrac{\partial f}{\partial \epsilon}(\epsilon,\mu) \mathcal{I}(\epsilon) d\epsilon,
\end{equation}
where $n_i$ is the ion density, $Z^*$ the mean ionic charge, $\mu$ the chemical potential and $f(\epsilon,\mu)$ the Fermi-Dirac distribution
\begin{equation}
f(\epsilon,\mu)=\dfrac{1}{e^{\beta(\epsilon-\mu)}+1},
\end{equation}
with $\beta=1/T$.

The function $\mathcal{I}(\epsilon)$ is related to the scattering cross-section $\Sigma(q)$ and to the ion-ion structure factor $S(q)$ by
\begin{equation}
\mathcal{I}(\epsilon)=\int_0^{2k}q^3 S(q) \Sigma(q) dq,
\end{equation}
where $\vec{q}=\vec{k}^\prime-\vec{k}$ is the momentum transferred in the elastic scattering event ({\it i.e.} in which $|\vec{k}^\prime|=|\vec{k}|$). Introducing the scattering angle $\theta\equiv (\vec{k},\vec{k}^\prime)$, one has $q^2=2k^2 (1-\chi)$, where $\chi=\cos\theta$, and one gets then the following expression introducing the squared modulus of the scattering amplitude $|a(k,\chi)|^2$
\begin{equation}
\mathcal{I}(\epsilon)=2k^4 \int_{-1}^1 S\left[k\sqrt{2(1-\chi)}\right]|a(k,\chi)|^2 (1-\chi) d\chi.
\end{equation}
$|a(k,\chi)|^2$ is provided by the $t-$matrix formalism of Evans \cite{evans1973} which reads, in the relativistic formalism underlying our average-atom code {\sc Paradisio} \cite{sterne2007}
\begin{equation}\label{scattering}
|a(k,\chi)|^2=\frac{1}{k^2}\left(\Big|\sum_\kappa |\kappa|e^{i\delta_\kappa(k)}\sin[\delta_\kappa(k)]P_{\ell}(\chi)\Big|^2+\Big|\sum_\kappa \frac{|\kappa|}{i\kappa}e^{i\delta_\kappa(k)}\sin[\delta_\kappa(k)]P^1_\ell(\chi)\Big|^2\right),
\end{equation}
where $\kappa=-(\ell+1)$ for $j=\ell+1/2$, $\kappa=\ell$ for $j=\ell-1/2$, $\ell$ being the usual orbital quantum number. $P_\ell$ and $P^1_\ell$ are the Legendre and associated Legendre polynomials. Finally, the quantities $\delta_\kappa(k)$ denote the scattering phase-shifts, provided by the average-atom code.\\

\subsection{Definition of mean ionic charge and corresponding chemical potential}
Mean ionic charge $Z^*$ is not a quantum-mechanical observable, and is therefore not clearly defined. Several definitions are possible in the framework of average-atom models. First, $Z^*$ can be identified with the ideally free part of the conduction electrons, whose number reads
\begin{equation}\label{Zfree}
Z_\mathrm{free}=\int_0^\infty f(\epsilon,\mu)X_\mathrm{ideal}(\epsilon) d\epsilon,
\end{equation}
where $X_\mathrm{ideal}(\epsilon)$ represents the ideal electron density of states
\begin{equation}
X_\mathrm{ideal}(\epsilon)=\dfrac{\sqrt{2\epsilon}}{\pi^2 n_i}.
\end{equation}
Actually, $Z_\mathrm{free}$ is related to the charge density at infinity. Within the framework of the {\sc Inferno} model, its value is given by $Z_\mathrm{free}=\overline{n}/n_i$, where $\overline{n}$ is the jellium density.\\ 

Definition (\ref{Zfree}), by considering only the charge density at infinity, excludes less extensive charge densities which can nevertheless contribute to the electrical conductivity. A second definition identifies $Z^*$ with the total number of continuum electrons
\begin{equation}\label{Zcont}
Z_\mathrm{cont}=\int_0^\infty f(\epsilon,\mu)X(\epsilon)d\epsilon,
\end{equation}
where $X(\epsilon)$ denotes the continuum density of states.\\

Unlike the first definition, the latter (\ref{Zcont}) suffers from the defect of including possible charges trapped in resonances, which do not contribute to the electrical properties \cite{sterne2007}. Other definitions for $Z^*$ can be introduced to try to overcome these pitfalls. They are inspired by Friedel's model of an impurity (here the ion) embedded in a perfect electron gas \cite{friedel1952}. Friedel's model expresses the modification of the uniform electron gas (UEG) charge by the impurity (I) as
\begin{equation}\label{def3}
Z^{\mathrm{I\,in\,UEG}}=Z^\mathrm{UEG}+Z_F,
\end{equation}
where $Z_F$ denotes the charge displaced by the electron-ion potential, whose value can be obtained applying the finite-temperature Friedel sum rule. In the framework of the relativistic formalism \cite{friedel1952,faussurier2021} one has
\begin{align}\label{ZF}
Z_F&=\dfrac{2}{\pi}\int_0^\infty d\epsilon f(\epsilon,\mu)\sum_\kappa |\kappa|\dfrac{\partial\delta_\kappa (\epsilon)}{\partial\epsilon}\nonumber\\ &=\dfrac{2}{\pi}\int_0^\infty d\epsilon \left(-\dfrac{\partial f}{\partial\epsilon}\right)\sum_\kappa |\kappa|\delta_\kappa (\epsilon).
\end{align}

The third definition that we will consider consists in adding the number of displaced charges to the definition (\ref{Zfree}), yielding
\begin{equation}\label{def5}
Z^*=Z_\mathrm{free}+Z_F=\dfrac{\overline{n}}{n_i}+Z_F.
\end{equation}

Eq.~(\ref{def3}) can be interpreted from a different point of view. Considering that $Z^{\mathrm{I\,in\,UEG}}=Z_\mathrm{cont}$, one gets $Z^\mathrm{UEG}=Z_\mathrm{cont}-Z_F$. A fourth possibility comes then from the electrical neutrality requirement that imposes $Z^\mathrm{UEG}=Z^*$:
\begin{equation}\label{def4}
Z^*=Z_\mathrm{cont}-Z_F.
\end{equation}
The latter definition follows the same spirit than the one $Z^*=Z_\mathrm{cont}-Z_\mathrm{quasi-b}$ proposed by Petrov and Davidson \cite{petrov2021}, following Sterne \textit{et al.}'s paper \cite{sterne2007}, for the case of dense plasmas when resonances appear in the continuum. $Z_\mathrm{quasi-b}$ denotes the number of electrons trapped in the resonances, considered as bound, despite their positive energy. Since the contributions of the resonances largely dominate those of the expanded free states in formula (\ref{ZF}), the latter appears appropriate for the calculation of $Z_\mathrm{quasi-b}$. The use of Eq.~(\ref{def4}) is therefore relevant at high densities.\\ 

Displaced charge $Z_F$ can be positive or negative, depending on the sign of the dominant $\delta_\kappa(\epsilon)$ phase-shifts. Negative phase shifts occur when the potential experienced by the electrons becomes more repulsive (see Ref.~\cite{messiah1961}, p.~405), {\it i.e.} when the attractive $-Z/r$ nucleus-electron contribution is more strongly counterbalanced by the repulsive Coulomb electron-electron contribution. Low densities $\rho$ increase the probability density of an electron in low $\ell$ states within the extended radius $R_\mathrm{WS}$, and thus favor the repulsive contributions to the potential. Equation~(\ref{def4}) implies in that case that some bound electrons (\textit{i.e.} of negative energy) must be considered as ``quasi-free'' and counted in the number of electrons contributing to the electrical conductivity. 
Equation~(\ref{def4}) appears then as an extension to the expanded plasma case of the formula proposed by Petrov and Davidson for the dense plasmas.\\ 

\begin{figure}
\begin{center}
\includegraphics[scale=0.40]{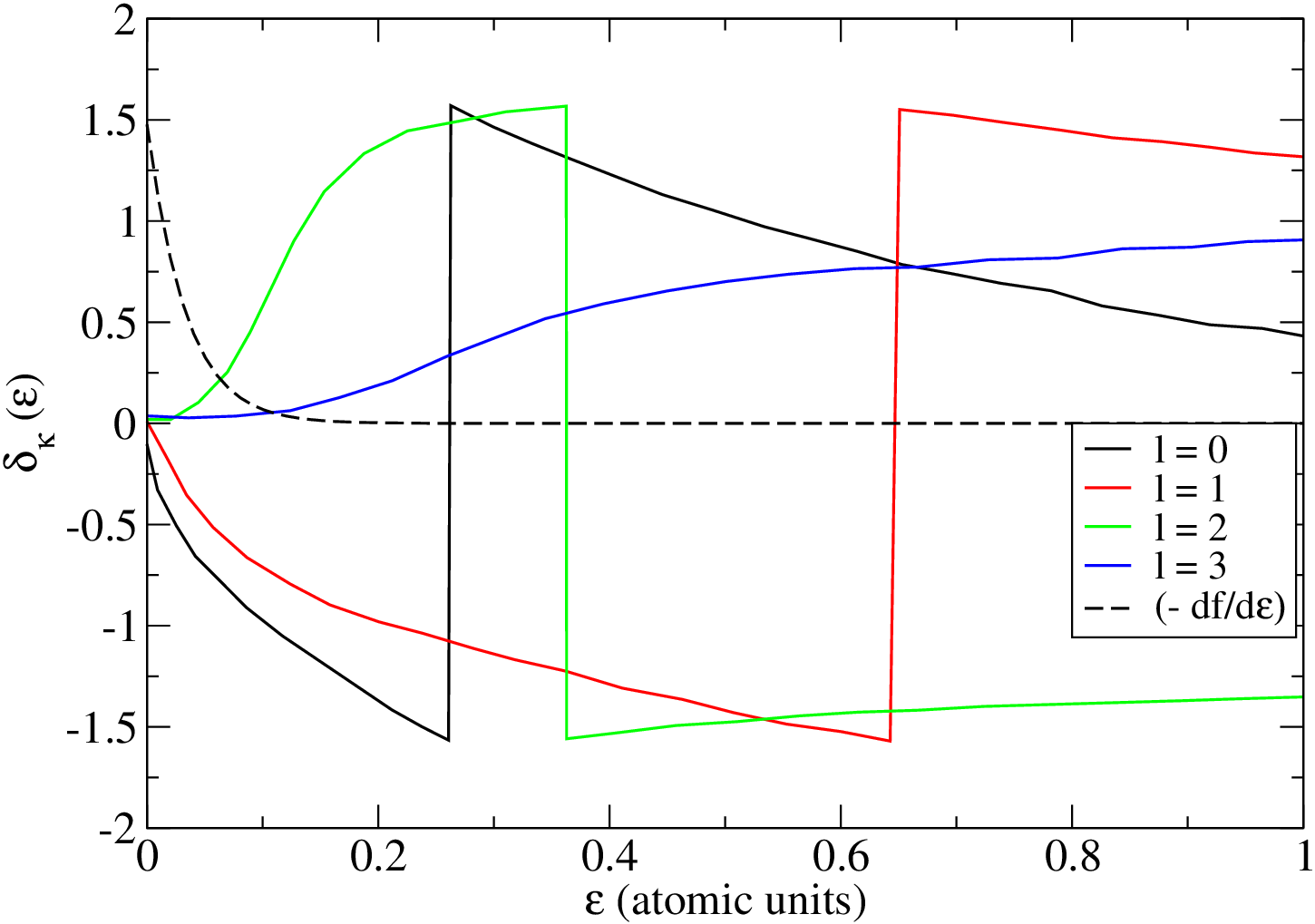}
\end{center}
\caption{\label{fig} Aluminum at $\rho=$ 0.3 g.cm$^{-3}$ and $T=$ 10 000 K: dominant phase-shifts ($\ell$=0, 1, 2 and 3), as functions of the energy. }
\end{figure}

Figure~\ref{fig} illustrates how $Z_F$ can be negative, by the example of aluminum at density $\rho=$ 0.3 g.cm$^{-3}$ and temperature $T=$ 10 000 K. The black, red, green and blue full lines represent phase-shifts obtained with {\sc Paradisio}. The dashed black line corresponds to $(-\partial f/\partial\epsilon)$ and underlines the dominant negative contributions of $\ell=0$ and 1 to $Z_F$. Therefore, Eq.~(\ref{def5}) subtracts ``s'' and ``p'' type states from an ideal electronic density of states, whereas Eq.~(\ref{def4}) introduces additional $\ell=0$ and $\ell=1$ ones to the average-atom conduction density of states. In the former case, the resulting $Z^*$ remains consistent with the ideal density of states assumed in Ziman's formalism, whereas, in the latter one, the inconsistency between the initial $Z^*=Z_\mathrm{cont}$ value with this assumption is reduced by the addition of extended ``s'' and ``p'' charges.\\

The Ziman formalism assumes that the scattered electrons are free. $Z^*$ is therefore associated to the chemical potential $\mu^*$ that verifies
\begin{equation}\label{mu_star}
Z^*=\int_0^\infty d\epsilon\, f(\epsilon,\mu^*) X_\mathrm{ideal}(\epsilon),
\end{equation}
or, equivalently
\begin{equation}\label{mu}
Z^*=\dfrac{\sqrt{2}}{\pi^2 n_i \beta^{3/2}}\mathcal{F}_{1/2}(\beta\mu^*),
\end{equation}
where we have introduced the Fermi function of order 1/2
\begin{equation}
\mathcal{F}_{1/2}(x)=\int_0^\infty dt \dfrac{t^{1/2}}{e^{t-x}+1}.
\end{equation}\\

The compatibility of $Z^*$ and the chemical potential $\mu^*$ obtained with Eq.~(\ref{mu_star}) has been discussed by Burrill {\it et al.} \cite{burrill2016}. The authors came to the conclusion that the use of total continuum charge $Z_\mathrm{cont}$ for the mean ion charge $Z^*$ should be associated to the exact average-atom chemical potential $\mu$, but this implies to account for the non ideality of the electron gas in Ziman's theory. However, in a review article \cite{ziman1970}, Ziman asserts that such a modification would have a limited impact on the resulting resistivities, due to compensations by the scattering contributions. The results presented in the next section were obtained with the original Ziman formalism based on the assumption of an ideal uniform electron gas. We will however (see section \ref{sec4}) examine this possibility in a case for which we observed the largest deviations from the ideal electron density of states, {\it i.e.} aluminum at $\rho=$ 0.3 g.cm$^{-3}$ and $T=$ 10 000 K.

\subsection{Quantities provided by {\sc Paradisio}}
These are the average ion charge $Z^*$ and electron-ion phase-shifts $\delta_{\kappa}(k)$. {\sc Paradisio} handles all electronic states (bound and continuum ones) on the same footing in a quantum-mechanical framework, essential condition to properly account for the effects of electron-electron interactions in the phase-shifts \cite{pain2010}. A detailed description of the code, including how the phase-shifts, $Z_\mathrm{free}$ and $Z_\mathrm{cont}$ are obtained, is given in the Appendix.

We studied, in Paper II, the impact of the exchange-correlation functional on the Ziman resistivity calculations in the case of solid density aluminum. We showed that finite-temperature functionals unquestionably improve agreement with experiments and QMD theories. Among the finite-temperature formulations probed in Paper II, the KSDT \cite{karasiev2014} and GDSMFB \cite{groth2017} formulations are both based on the same Pad\'e approximants, involving a set of 18 parameters. Whereas KSDT obtained them by fitting  Restricted Path Integral Monte-Carlo (RPIMC) data, GDSMFB exclusively used Configuration Path Integral Monte-Carlo and Blocking Path Integral Monte-Carlo calculations. The advantage of the latter is that they do not invoke the fixed node approximation which is the basis of RPIMC calculations, and may limit their relevance. GDSMFB is therefore more accurate than KSDT, although the differences with KSDT can be small in some phase space regions of the UEG. In the present work we retain these latter GDSMFB functionals, although this choice did not change significantly our results compared to the ones obtained with KSDT.

\subsection{The sensitivity of Ziman's formalism to the ion-ion structure factor}
For the sake of consistency with Paper I, the Ornstein-Zernike equation is solved together with the Hypernetted-chain (HNC) closure relation for a system of screened charged spheres, following Rogers \cite{rogers1980}. The calculations are initiated using the model direct correlation function of Held and Pignolet \cite{held1986}. Following Paper I, elastic scattering contributions are then removed from the HNC $S(q)$, which is equivalent to add the following quantity to Ziman's resistivity \cite{wetta2020}
\begin{equation}
\delta\eta=-\dfrac{1}{3\pi {Z^*}^2 n_i}\sum_G \dfrac{N(G)}{4\pi}\mathrm{e}^{-2W(G)}
\times \int_{G/2}^\infty \left(-\dfrac{\partial f}{\partial k} \right) k^2 G^2 \times\Bigg|a\left(k,1-\dfrac{G^2}{2k^2}\right)\Bigg|^2 dk,
\end{equation}
where $N(G)$ denotes the number of reciprocal lattice vectors of same length $G$ and $\mathrm{e}^{-2W(G)}$ the Debye-Waller factors accounting for thermal decay of the elastic contributions. In the thermodynamic conditions of the present work, we found that $\delta\eta\approx 0$. There are two reasons for that. First, temperatures are high enough to strongly attenuate the $G$'s contributions by the means of the Debye-Waller factors. Second, $\left(-\partial f/\partial\epsilon \right)=-k \left(\partial f/\partial k \right)$ is a peak centered at the chemical potential $\mu^*$, which is significantly located below zero at the considered temperatures and densities, so that the reciprocal vectors $G$ lie within the tail of $\left(-\partial f/\partial\epsilon \right)$.

In Paper II, we showed, in the case of solid density aluminum at 1 eV, the impact of $S(q)$ on the Ziman-Evans resistivity, and observed that it was rather limited, despite strong discrepancies between the models we probed. Here also, the $\left(-\partial f/\partial\epsilon \right)$ function plays a determinant role by strongly attenuating the contributions of the high energies, for which the various probed $S(q)$ differ the most. In the present work, the ion coupling parameter $\Gamma=\left(Z^* e \right)^2/(R_\mathrm{WS}\,k_BT)$, where $R_\mathrm{WS}$ is the Wigner-Seitz cell radius (see Eq.~(\ref{Rws}) in the appendix), is also much lower than in the conditions of Paper II, reducing the discrepancy between the HNC structure factor and the more appropriate to the liquid state one-component-plasma (OCP) one.

This surprising low sensitivity of Ziman's resistivity to $S(q)$ reflects the fact that, when {\sc Inferno}-type average-atom codes are used, $S(q)$ only impacts the resistivity by the means of ZE formula's explicit dependence on this quantity. At the opposite, more sophisticated average-atom methods that go beyond the jellium approximation, like neutral-pseudo-atom based ones \cite{dharma2021a,dharma2021b,stanek2021}, allow self-consistent calculation of the ion-ion structure factor and electronic states. $S(q)$ then also impacts $\eta$ indirectly, by the means of the scattering phase-shifts $\delta_\kappa$, as well as by $Z^*$. 

\section{Comparisons to experimental results and other theoretical models}\label{sec3}

\begin{figure*}
    
     \begin{subfigure}[1]{0.45\textwidth}
         \centering
         \includegraphics[width=\textwidth]{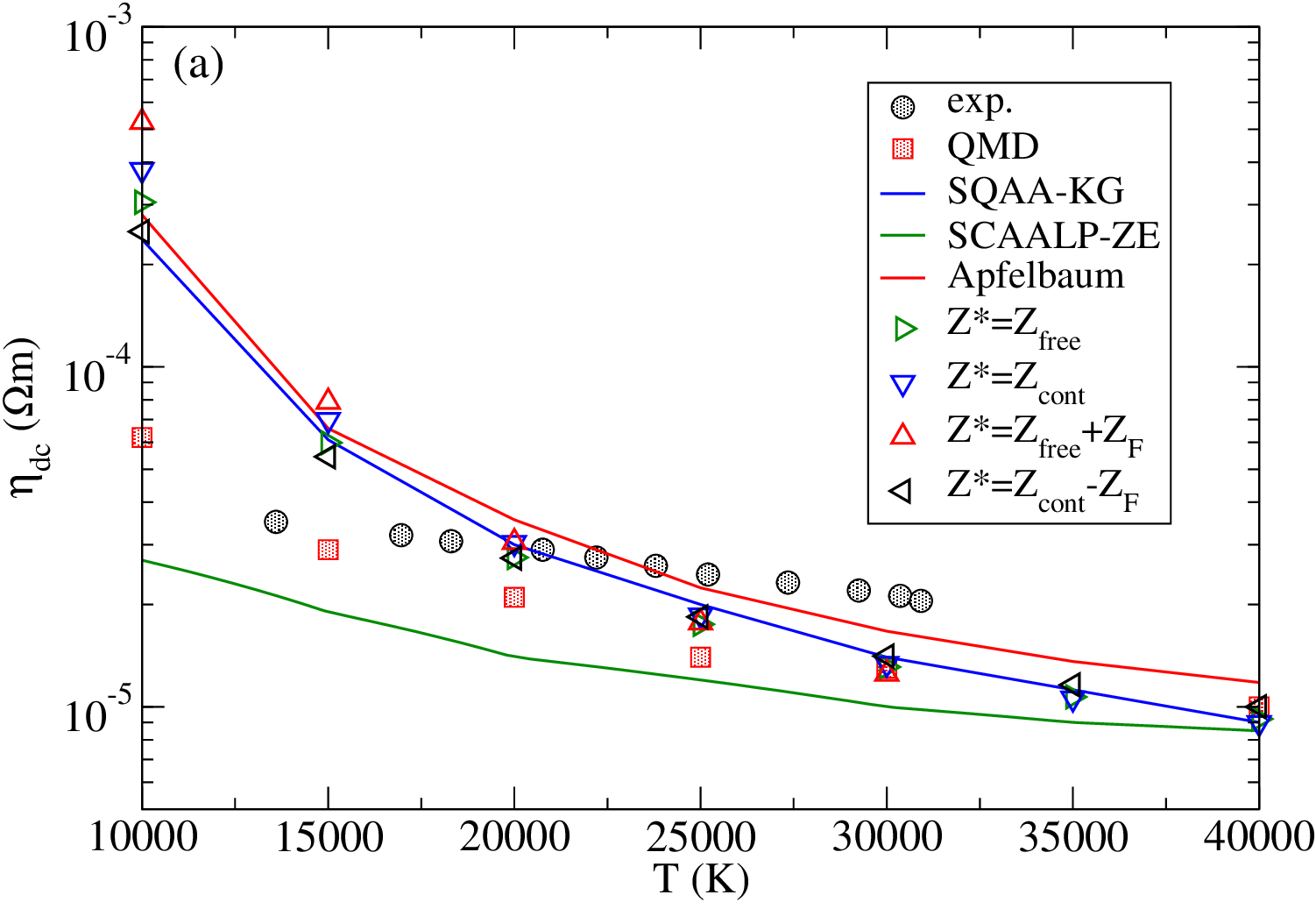}
     \end{subfigure}
     \hfill
     \begin{subfigure}[2]{0.45\textwidth}
         \centering
         \includegraphics[width=\textwidth]{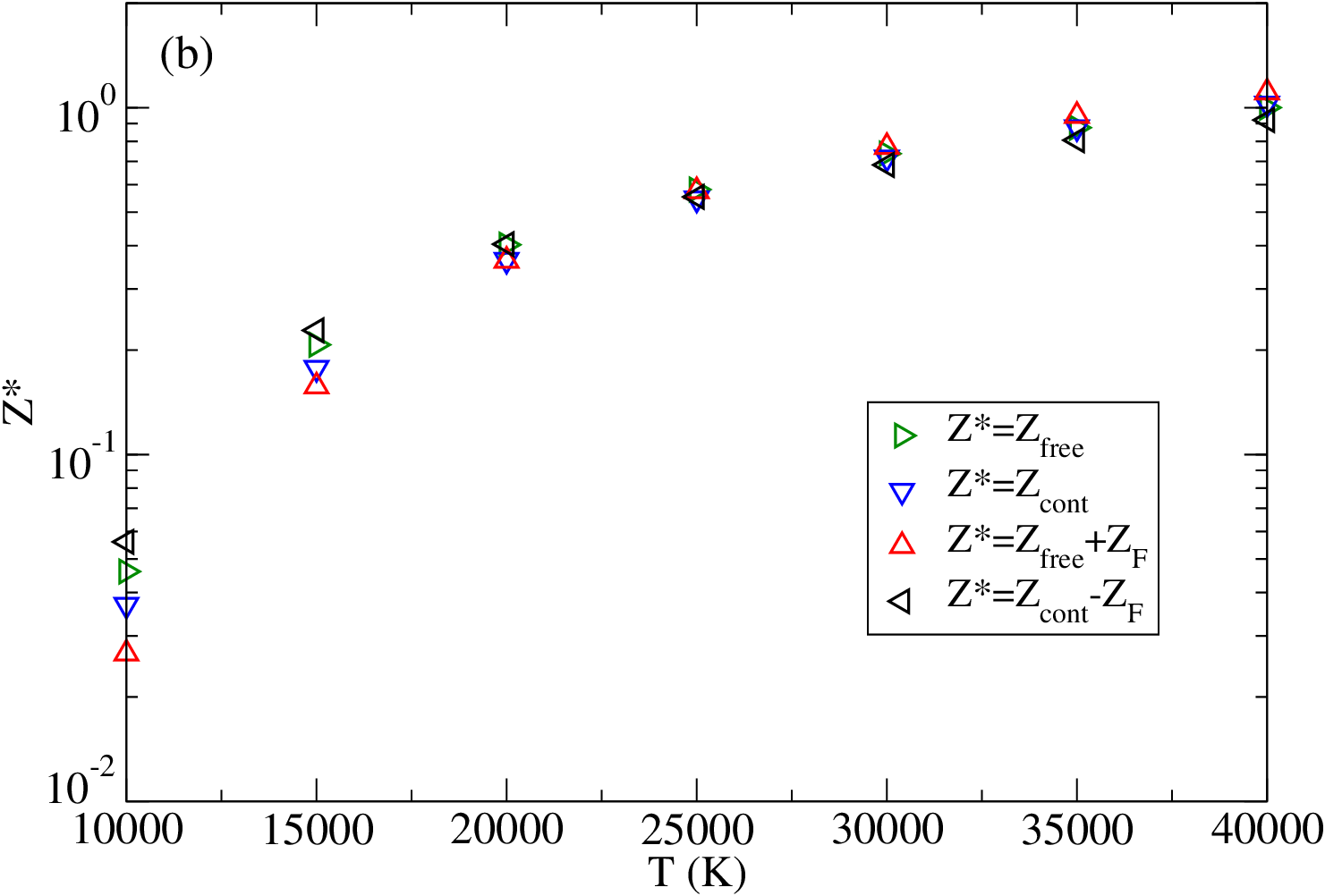}
     \end{subfigure}
     
\caption{Titanium at $\rho=$ 0.2 g.cm$^{-3}$ ($n_i=2.5\ 10^{21}$ cm$^{-3}$). (a): Experiments (filled gray circles), QMD calculations (filled red squares), Kubo-Greenwood approach using SQAA results (blue line), and the Ziman-Evans one using the quantities calculated by SCAALP (green line) are all from reference \cite{clerouin2012}. The red line corresponds to calculations based on a chemical model \cite{apfelbaum2017}. Our own results (Ziman-Evans formulation with quantities given by {\sc Paradisio}) are represented by the triangles: green one for $Z^*=Z_\mathrm{free}$, blue one assuming $Z^*=Z_\mathrm{cont}$ and red triangles based on the definition $Z^*=Z_\mathrm{free}+Z_F$. (b): Values of $Z_\mathrm{free}$, $Z_\mathrm{cont}$, $(Z_\mathrm{free}+Z_F)$ and $(Z_\mathrm{cont}-Z_F)$ along the 0.2 g.cm$^{-3}$ isochore.} 
\label{fig1}
\end{figure*}

\begin{figure*}

     \begin{subfigure}[1]{0.45\textwidth}
         \centering
         \includegraphics[width=\textwidth]{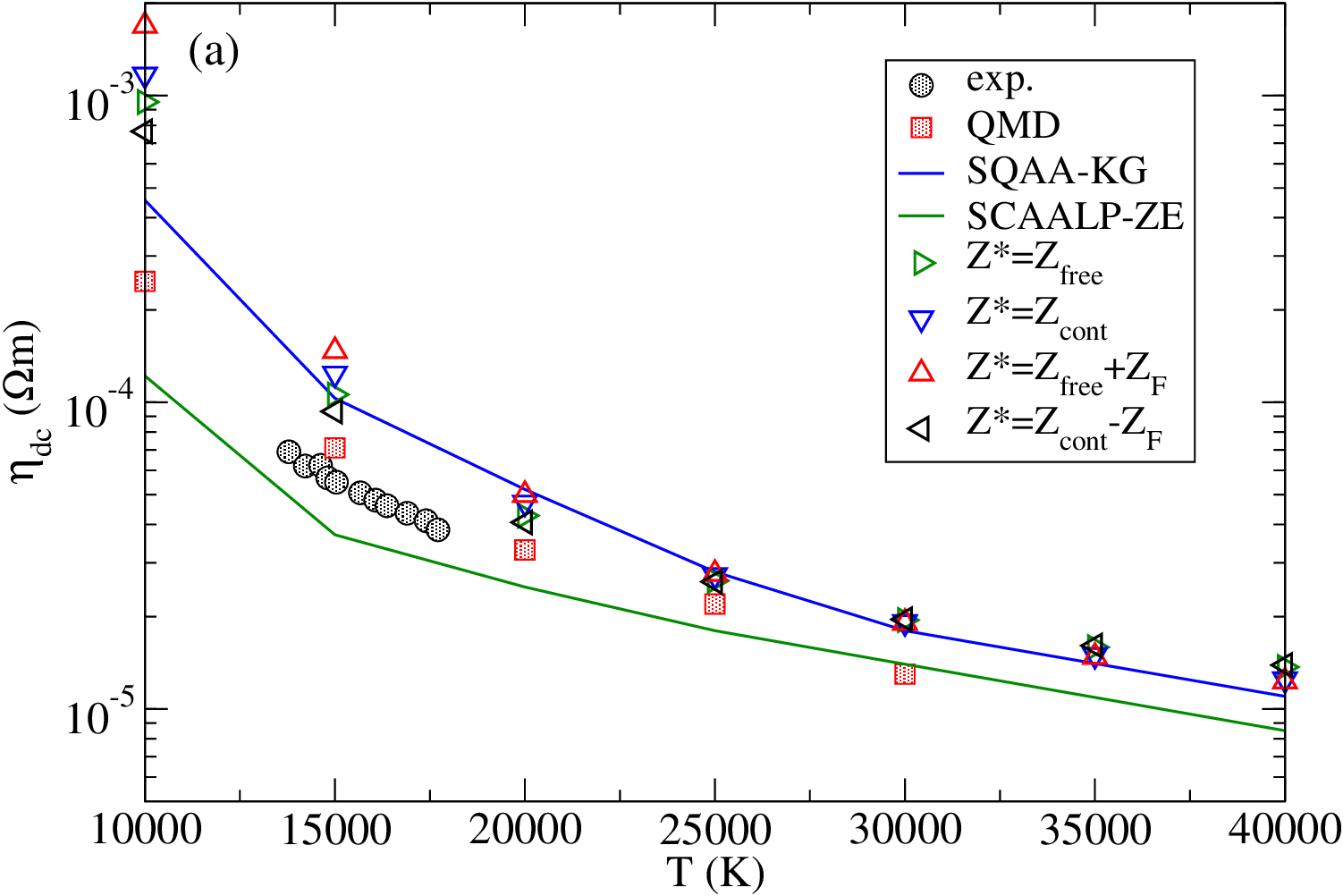}
     \end{subfigure}
     \hfill
     \begin{subfigure}[2]{0.45\textwidth}
         \centering
         \includegraphics[width=\textwidth]{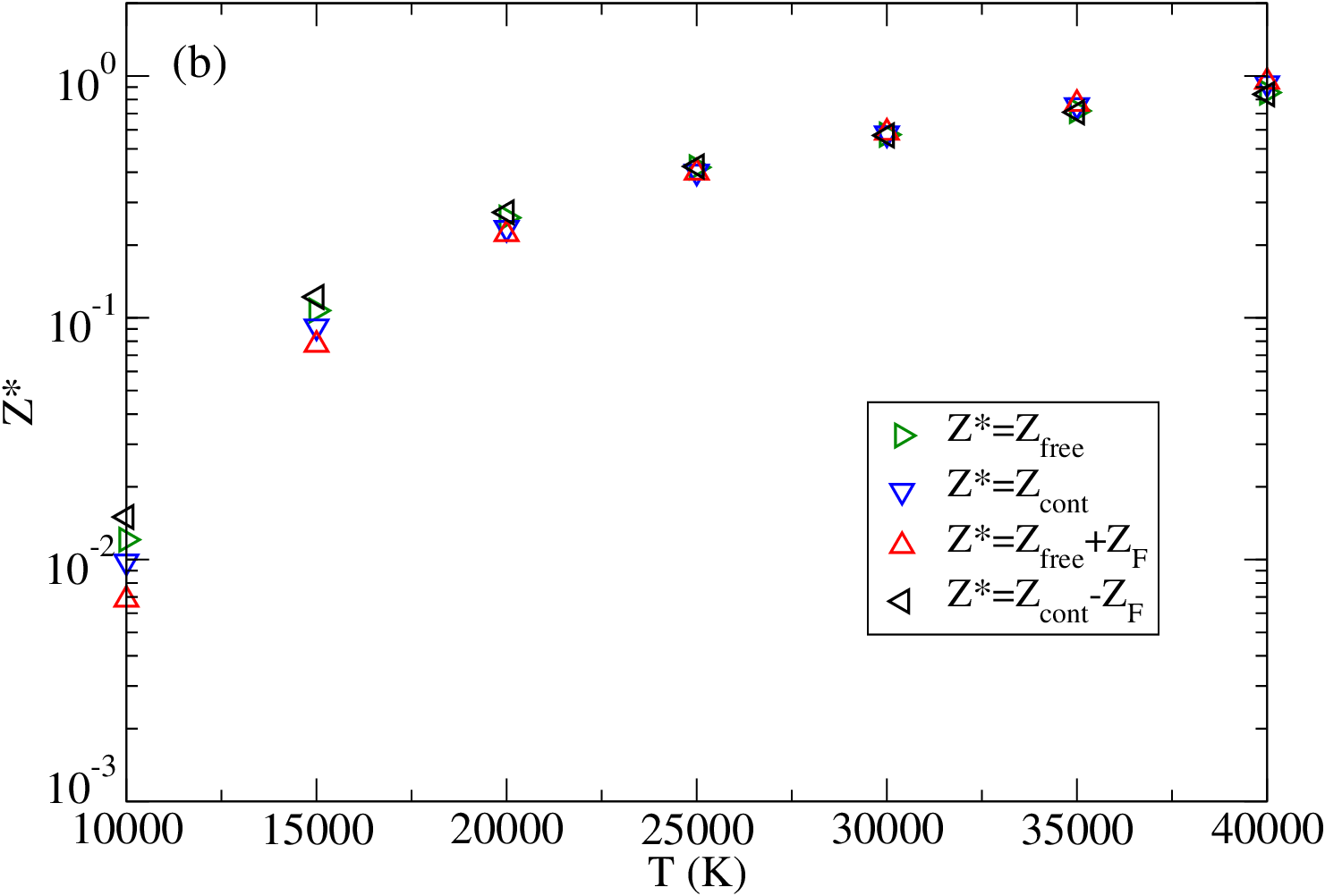}
     \end{subfigure}

\caption{Copper at $\rho=$ 0.3 g.cm$^{-3}$ ($n_i=2.8\ 10^{21}$ cm$^{-3}$). Same legends as for Fig.~\ref{fig1}.} 
\label{fig2}
\end{figure*}

\begin{figure*}

     \begin{subfigure}[1]{0.45\textwidth}
         \centering
         \includegraphics[width=\textwidth]{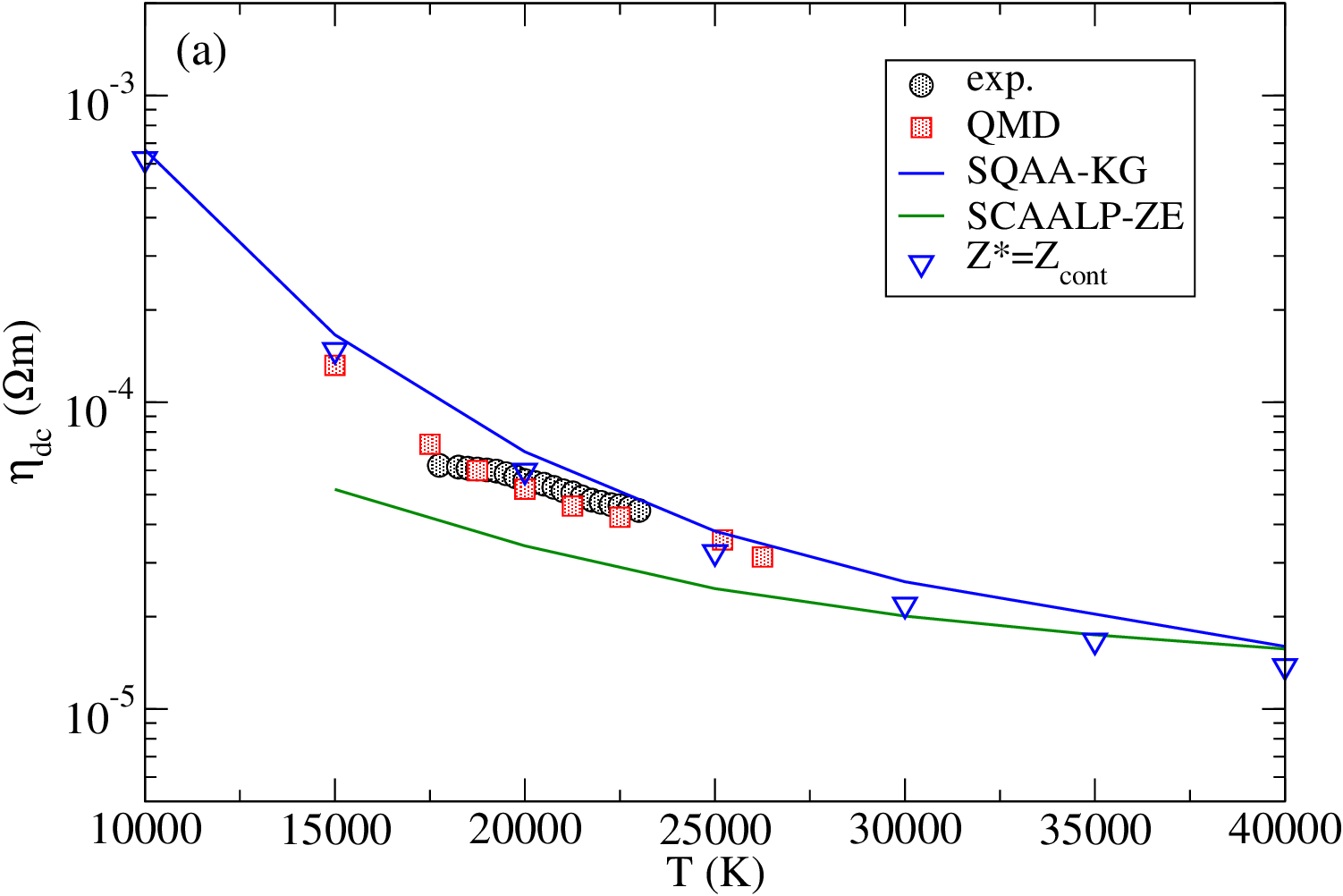}
     \end{subfigure}
     \hfill
     \begin{subfigure}[2]{0.45\textwidth}
         \centering
         \includegraphics[width=\textwidth]{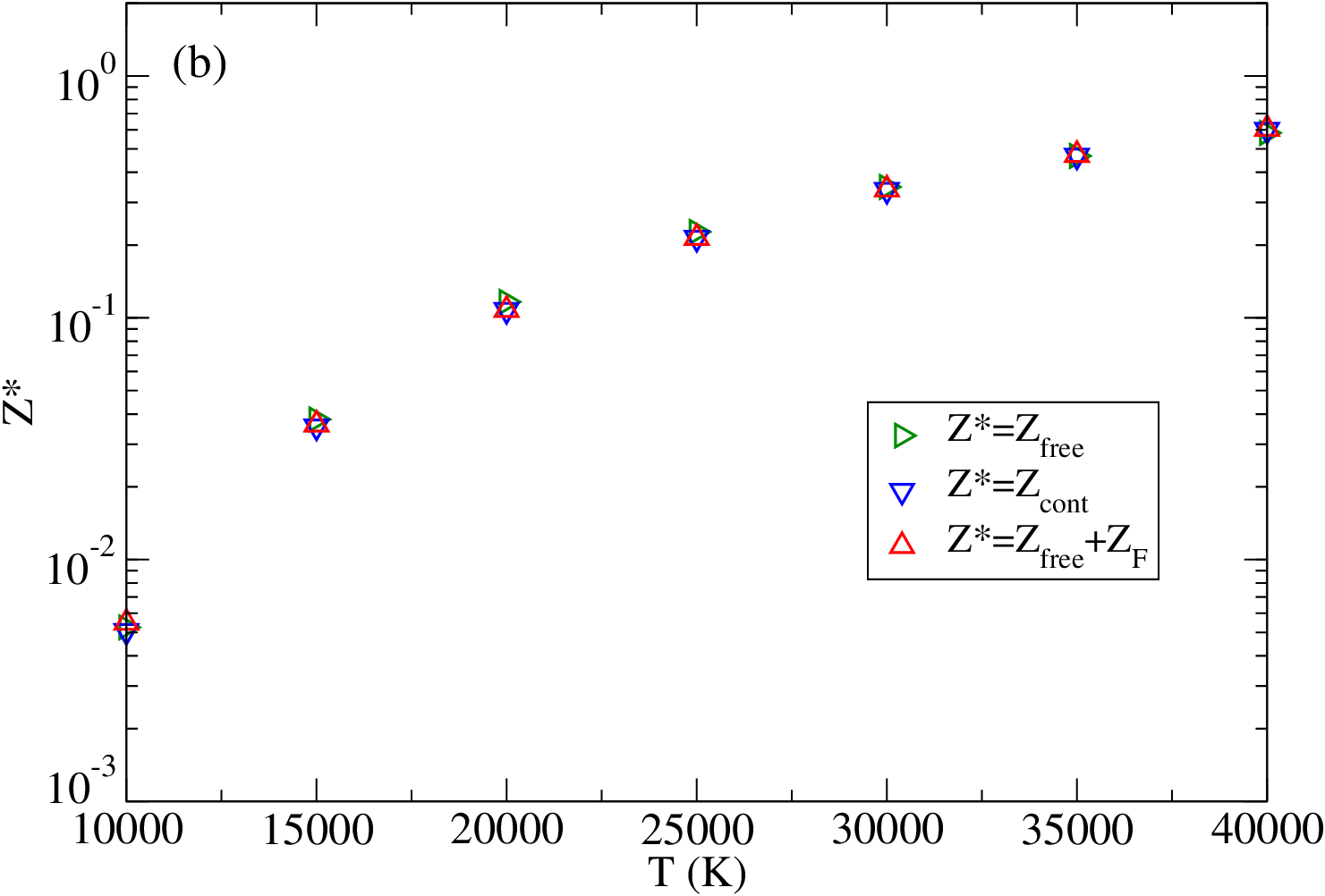}
     \end{subfigure}

\caption{Boron at $\rho=$ 0.1 g.cm$^{-3}$ ($n_i=5.2\ 10^{21}$ cm$^{-3}$). Same legends as for Fig.~\ref{fig1}.} 
\label{fig3}
\end{figure*}

\begin{figure*}

     \begin{subfigure}[1]{0.45\textwidth}
         \centering
         \includegraphics[width=\textwidth]{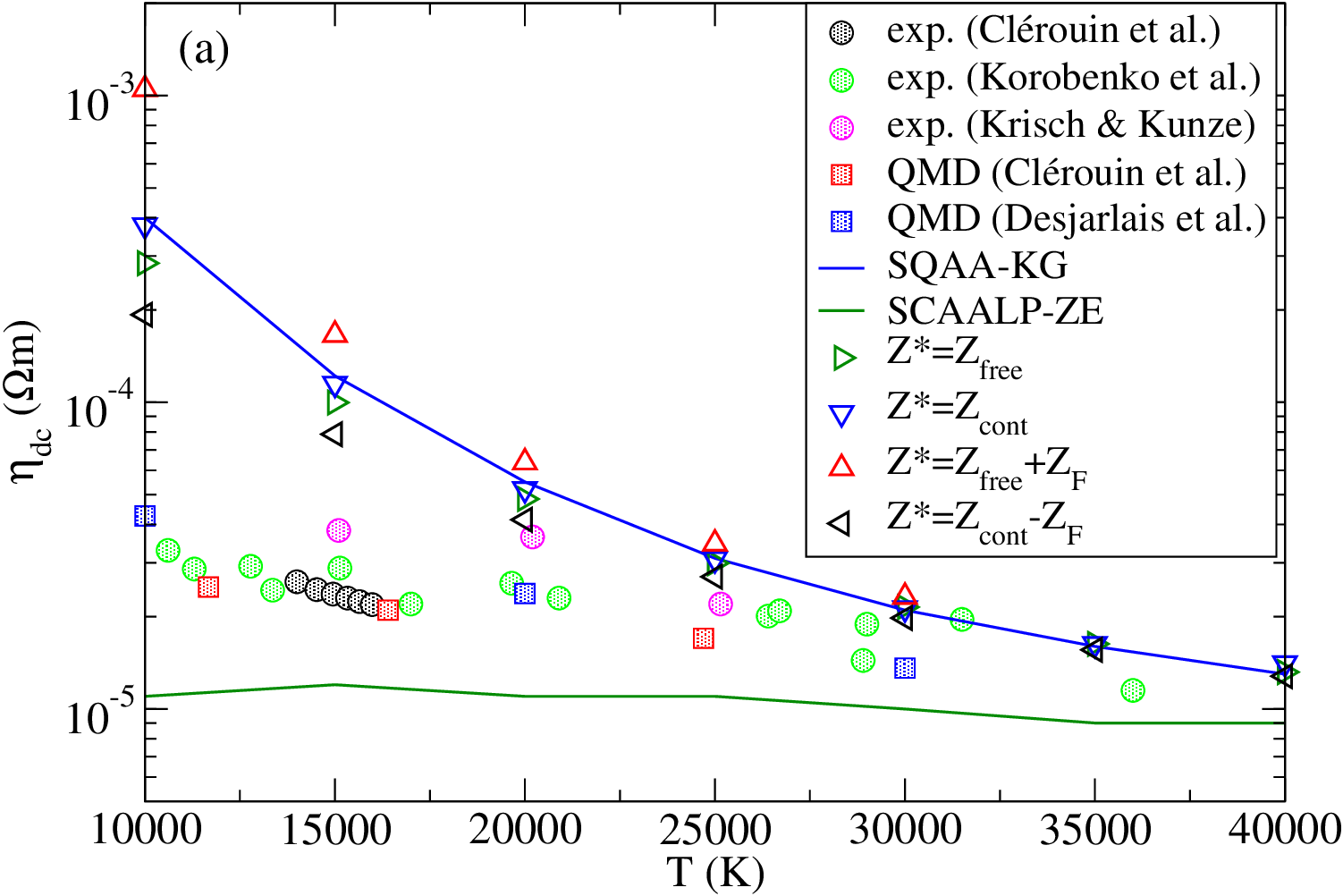}
     \end{subfigure}
     \hfill
     \begin{subfigure}[2]{0.45\textwidth}
         \centering
         \includegraphics[width=\textwidth]{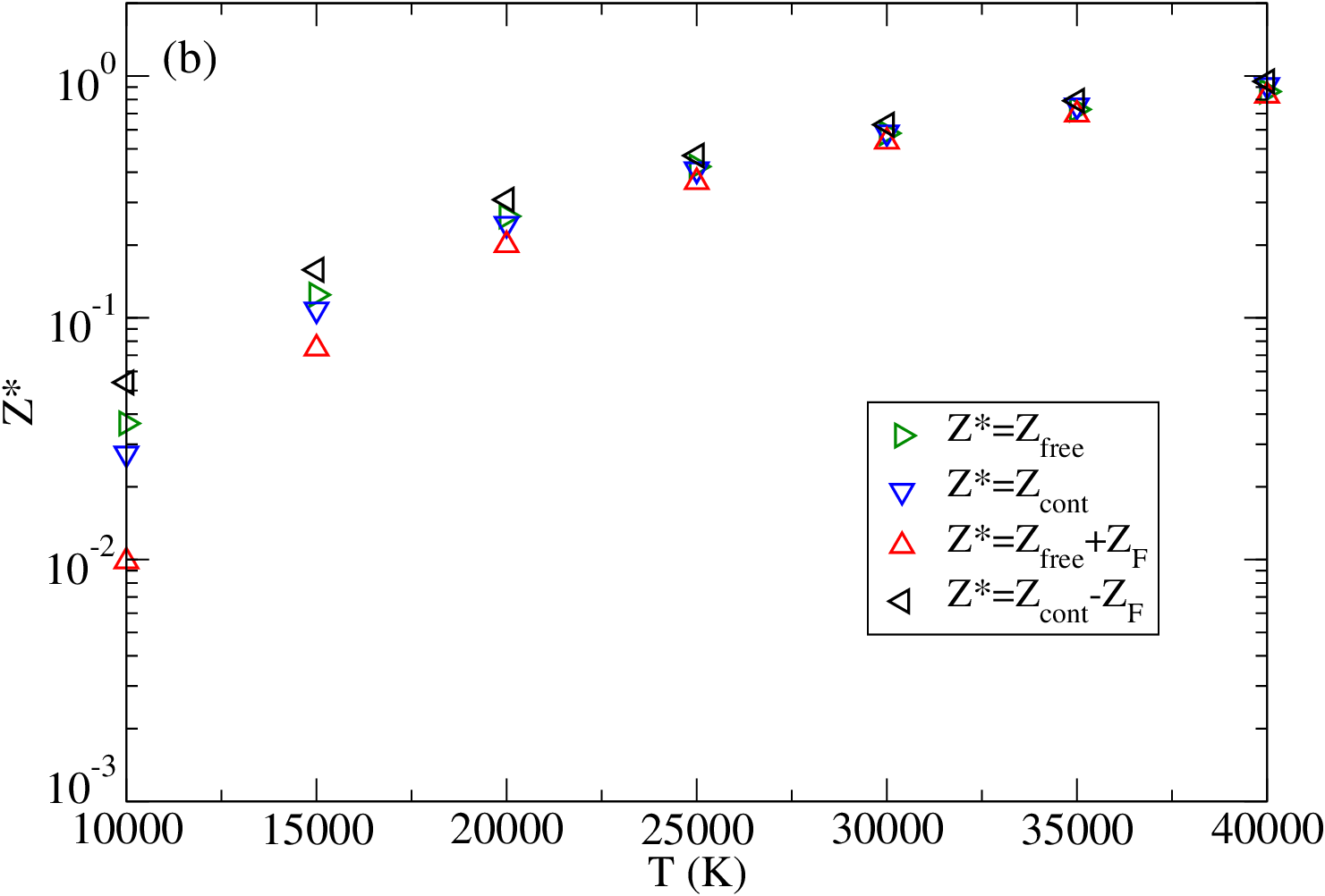}
     \end{subfigure}

\caption{Aluminum at $\rho=$ 0.3 g.cm$^{-3}$ ($n_i=6.7\ 10^{21}$ cm$^{-3}$). (a): In addition to Cl\'erouin {\it et al.}'s experiments (filled gray circles), the figure presents those of Korobenko {\it et al.} (filled green circles) \cite{korobenko2005} as well as measurements of Krisch and Kuntze \cite{krisch1998} (filled magenta circles). The theoretical data set is also enriched by Desjarlais {\it et al.}'s results (filled blue squares). (b): Values of $Z_\mathrm{free}$, $Z_\mathrm{cont}$, $(Z_\mathrm{free}+Z_F)$ and $(Z_\mathrm{cont}-Z_F)$ along the 0.3 g.cm$^{-3}$ isochore.}
\label{fig4}
\end{figure*}

We compare our calculations to the experimental electrical resistivities presented in Cl\'erouin {\it et al.}'s compilation of experiments performed on the ``isochoric plasma closed vessel'' facility \cite{clerouin2012}. This paper also reports the co-authors' theoretical results, in this case, quantum-molecular-dynamics and average-atom ones.\\
Among the elements selected in the present work, aluminum has been the subject of a great number of studies, both experimental and theoretical. We add comparisons to experiments performed by Korobenko {\it et al.} \cite{korobenko2005} and by Krisch and Kunze \cite{krisch1998}. The theoretical set retained for aluminum is enriched by the quantum-molecular-dynamics results of Desjarlais {\it et al.} \cite{desjarlais2002}.\\
Cl\'erouin {\it et al.}'s quantum-molecular-dynamics simulations are performed using projector augmented wave (PAW) pseudo-potentials with the electronic structure package VASP (Vienna Ab initio Simulation Package) developed at university of Vienna \cite{Kresse1993}, in the isokinetic ensemble in order to ensure a control of temperature. The exchange-correlation contribution is treated in the local density approximation (LDA) using the Ceperley-Alder parametrization \cite{Ceperley1980} in the case of boron and aluminum, and, for titanium and copper, in the generalized gradient approximation (GGA) using the Perdew-Zunger formulation \cite{Wang1991}. The resistivities are computed within the Kubo-Greenwood (KG) formalism \cite{Kubo1957,Greenwood1958} using the QMD Kohn-Sham orbitals $\psi_n^{\vec{k}}$, energies $\epsilon_n$ and occupations $f_n$ obtained for selected configurations: 
\begin{equation}
\sigma(\omega)=\frac{2\pi}{3\omega}n_i\sum_{n,m,\alpha}\sum_{\Vec{k}}W(\Vec{k})(f_n-f_m)\left|\langle\psi_n^{\Vec{k}}|\nabla_{\alpha}|\psi_m^{\Vec{k}}\rangle\right|^2\delta\left(\epsilon_m-\epsilon_n-\hbar\omega\right).
\end{equation}
$\Vec{k}$ and $W(\Vec{k})$ are the wave vectors and their weights in the Brillouin zone, $\nabla_{\alpha}$ ($\alpha=x,y,z$) the velocity operator in each direction between two states $n$ and $m$ with occupation $f_n$ and $f_m$.\\
General good agreement between quantum-molecular-dynamics theoretical results, represented by the filled red squares throughout Figs.~\ref{fig1} to \ref{fig4}, and experiments (filled gray circles) is observed.\\
The average-atom theoretical results presented in Ref.~\cite{clerouin2012} are based on two types of models. SQAA (Standard Quantum Average-Atom) \cite{clerouin2012CPP} is an {\sc Inferno} type average-atom code, similar to the {\sc Purgatorio} and {\sc Paradisio} ones. The specificity of the {\sc Inferno} model is to introduce the density effects through the boundary conditions at the edge of the ion sphere, with a uniform jellium replacing the ionic environment. The SCAALP (Self Consistent
Approach for Astrophysical and Laboratory Plasmas) code \cite{Blancard2004} goes beyond the jellium approximation by taking into account the ionic structure through a variational approach based on the Gibbs-Bogolyubov inequality for hard spheres. The electronic and ionic properties are then determined in a self-consistent way.\\
The resistivities calculated in Ref.~\cite{clerouin2012} are obtained by applying the Kubo-Greenwood method to the SQAA results, while the Ziman-Evans formalism is used with those from SCAALP. They are respectively represented by the blue and green curves. Both studies predict comparable resistivities at $T\gtrsim$ 30 000 K, but SQAA departs significantly from QMD calculations and from the experimental values at $T\lesssim$ 20 000 K, at the opposite of the SCAALP results, which show at least qualitative agreement with experiment, with less steep rises of the resistivities as $T$ decreases.\\
Our own calculations combine the Ziman-Evans formalism with the {\sc Inferno} based {\sc Paradisio} code, thus providing a third average-atom approach. They are represented by the triangles, and obtained with the four definitions (\ref{Zfree}), (\ref{Zcont}), (\ref{def3}) and (\ref{def4}) for the ion charge $Z^*$ proposed in the preceding section. The figures are sorted following increasing ion density $n_i$.

Subfigures (b) present, for each element considered in this paper, the values of the mean ion charge obtained with the definitions proposed in section \ref{sec2}. Except for the case of boron of density $\rho=$ 0.1 g.cm$^{-3}$, the four values of $Z^*$ diverge as temperature drops under 20 000 K typically, reflecting growing deviations from the UEG. Subfigures (a) show the calculated resistivities using the $Z^*$ values from subfigures (b). It appears that our {\sc Paradisio}-ZE results are in agreement with those of SQAA-KG in all situations where a single value of the average ionization can be obtained using {\sc Paradisio}, {\it i.e.} typically above 20 000 K and over the whole temperature range in the case of boron. Below 20 000 K, one observes, for the highest ion densities considered (0.3 g.cm$^{-3}$ for aluminum and 0.1 g.cm$^{-3}$ for boron), the best agreement with SQAA-KG when the definition $Z^*=Z_\mathrm{cont}$ is applied. For the two lowest ion densities (titanium at 0.2 g.cm$^{-3}$ and copper at 0.3 g.cm$^{-3}$), the resistivities calculated with the fourth definition (\ref{def4}) are the closest to SQAA-KG one at $T=$ 10 000 K. But for all considered ion densities, the use of the $Z^*=Z_\mathrm{cont}-Z_F$ improves the agreement between our {\sc Paradisio}-ZE calculations and experiments.\\ 

As mentioned above, the case of boron at 0.1 g.cm$^{-3}$ presents, among the ones considered in the present work, the best agreement with experiments and QMD calculations. Figure \ref{fig3}(b) shows that $Z_\mathrm{free}\approx Z_\mathrm{free}+Z_F\approx Z_\mathrm{cont}\approx Z_\mathrm{cont}-Z_F$ (the last equality follows from the preceding ones) on the whole investigated temperature range 10 000 K$\leq T\leq$ 40 000 K. The first equality implies that $Z_F\approx 0$, which  reflects the fact that electrons are clearly separated into perfectly bound and perfectly free ones. Moreover, we can conclude from $Z_\mathrm{free}\approx Z_\mathrm{cont}$ that all continuum electrons form an ideal uniform  electron gas. These are the best conditions for the application of Ziman's theory to electrical resitivity calculations. Since there is no ambiguity  about which electrons contribute or not to conductivity, the AA models provide a unique value for the mean ion charge, and Ziman's resistivity is unique too. $Z^*$ is also perfectly consistent with the UEG DOS assumption which underpins the Ziman theory. The case of boron at 0.1 g.cm$^{-3}$ shows that, when these two requirements are fulfilled, good agreement between AA-ZE and QMD calculations is achievable.

As concerns titanium, copper and aluminum, if we focus on temperatures below 15 000 K, we observe, based on figures \ref{fig1} up to \ref{fig4}, a hierarchy (not a rule) $\eta^\mathrm{exp.}< \eta(Z_\mathrm{cont}-Z_F)< \eta(Z_\mathrm{free})< \eta(Z_\mathrm{cont})< \eta(Z_\mathrm{free}+Z_F)$ between the resistivities $\eta(Z^*)$ obtained with the different definitions for the mean ionic charge. At least for the few materials considered in this work, the best agreement with the experimental values $\eta^\mathrm{exp.}$ is obtained with the two definitions $Z^*=Z_\mathrm{cont}-Z_F$ and $Z^*=Z_\mathrm{free}$ which are the most consistent with the description of the scattered electron gas as a perfect one. This is obvious when $Z^*$ is identified to the jellium's charge. According to figures \ref{fig1} up to \ref{fig4}, our {\sc Paradisio}-ZE results are improved when $Z^*$ is identified to the charge deduced from a density of states obtained by removing non-ideal contributions from the average-atom continuum density of states, thus approaching a ``quasi-free like'' density of states. At the opposite, the two definitions $Z^*=Z_\mathrm{cont}$ and $Z^*=Z_\mathrm{free}+Z_F$ show the worst agreement with experiments. The first case takes $Z^*$ as the number of electrons obtained from the total continuum density of states, whereas the latter identifies it to the number contained in a ``real like'' density of states, {\it i.e.} the ideal one perturbed by the presence of the ion.\\

\begin{figure}
\centering
\vspace{1cm}
\includegraphics[scale=0.40]{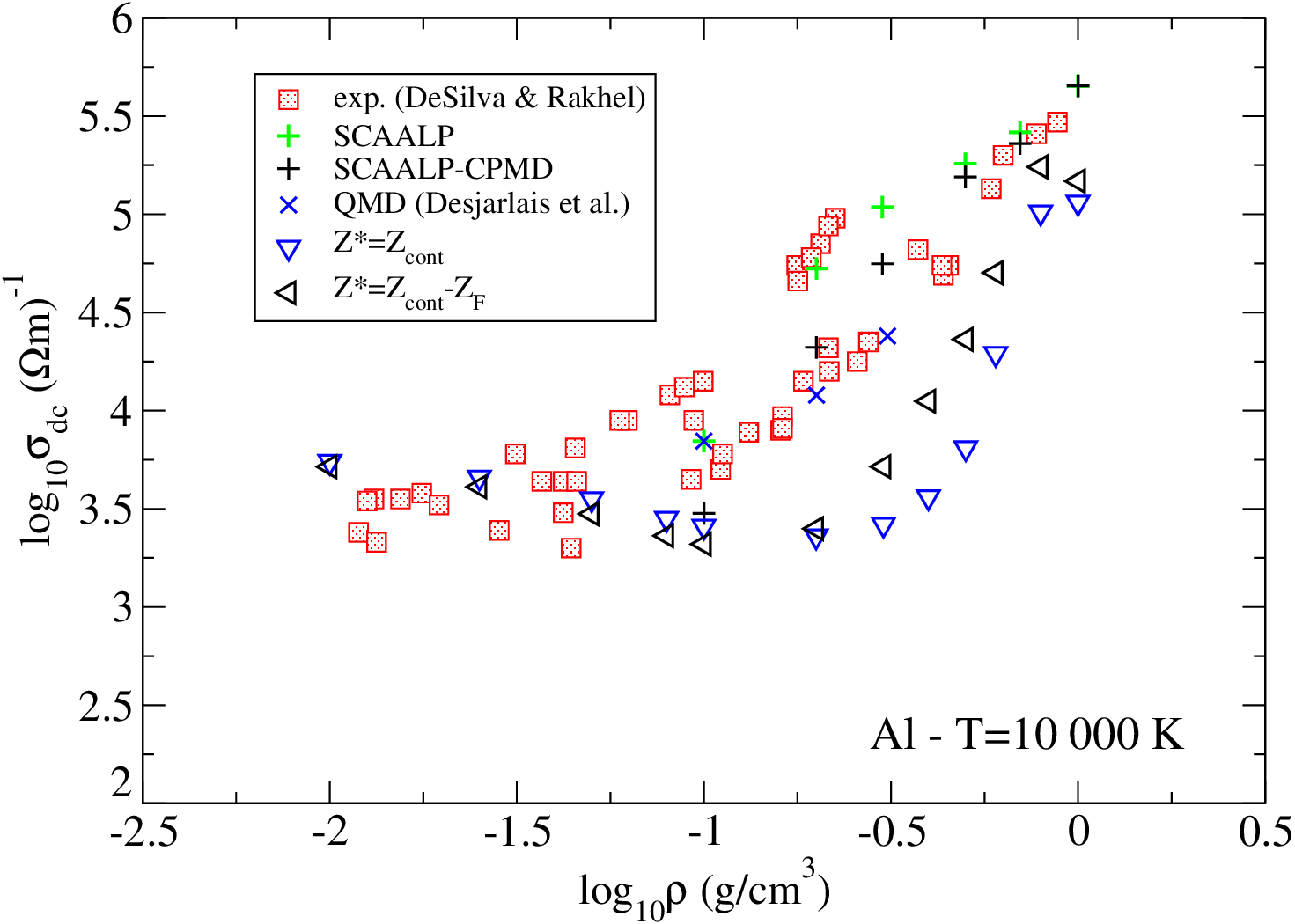}
\caption{\label{fig5}Aluminum's electrical conductivity $\sigma_\mathrm{dc}$ along the $T=$ 10 000 K isotherm in the density range of metal to non-metal transition. Red squares: experiments \cite{desilva2005}. Green pluses: SCAALP calculations \cite{faussurier2006JQSRT,faussurier2006PRB}. Black pluses: from the same authors, results obtained with a combination between SCAALP and Car-Parrinello Molecular-Dynamics (CPMD). Blue crosses: Desjarlais {\it et al.}'s QMD results \cite{desjarlais2002}. Triangles: our calculations, in blue with $Z^*=Z_\mathrm{cont}$, and in black with $Z^*=Z_\mathrm{cont}-Z_F$.}
\end{figure}

Among the few examples studied in the present work, we observe the greatest differences in the mean ionization $Z^*$ when different definitions are used, for aluminum at $\rho=0.3$ g/cm$^3$ and at $T=$ 10 000 K. More specifically, we obtain in that case $Z_\mathrm{cont}-Z_F\approx \frac{5}{3} Z_\mathrm{cont}$, {\it i.e.} $Z_F\approx -\frac{2}{3} Z_\mathrm{cont}$. DeSilva and Rakhel \cite{desilva2005} measured electrical conductivities for a few metals at densities covering the metal to non-metal transition. They performed a great number of experiments on aluminum at $T=$ 10 000 K, and obtained a rather dispersed cloud of experimental values for densities ranging from 0.01 g.cm$^{-3}$ up to 1 g.cm$^{-3}$, represented in figure \ref{fig5} by the red squares. According to our own dispersion of calculated conductivities at 10 000 K and at 0.3 g.cm$^{-3}$, it appeared to us interesting to extend our calculations for aluminum in the density range covered by DeSilva and Rakhel, in order to verify if our values obtained with definition $Z^*=Z_\mathrm{cont}-Z_F$ could enter into the experimental cloud represented by the red squares. As in figures \ref{fig1} up to \ref{fig4}, the blue triangles of figure \ref{fig5} correspond to our calculations with $Z^*=Z_\mathrm{cont}$ and the black ones to those obtained with $Z^*=Z_\mathrm{cont}-Z_F$. We can observe that our {\sc Paradisio}-ZE values are improved with the latter definition for the mean ionization, but not enough to bring them within the cloud of experimental values.\\

Of course, other definitions can be considered, and likely to improve the agreement with experiments. While we were carrying out the present work, Callow {\it et al.} published new ways to calculate mean ionization in the framework of average-atom models \cite{callow2023}, consisting in partitioning bound and free electrons according to their shells (characterized by principal quantum number $n$) instead of the sign of their energies. But the arbitrary character of the assignment of electrons according to their energies is then replaced by the necessity to clearly set the boundary between bound and free orbitals. These news methods are therefore not self-contained ones, in contrast to those we proposed in the present work. Indeed, all the definitions we retained for the mean ionization can be provided by {\sc Paradisio} which only requires the atomic number, the molar mass, a mass density and a temperature as input parameters (see the appendix). This does not call into question the quality of Callow {\it et al.}'s work. We however believe that it is not certain that the methods they propose for the calculation of the average ionization can be applied to extended temperature and/or density domains without any readjustment. Moreover, we are in the present work seeking for a definition for $Z^*$ consistent with Ziman's approach for the electrical resistivity, whereas Callow {\it et al.} place themselves in a more general framework, aiming at proposing a global definition applicable to the greatest number of numerical methods involved in the study of the properties of the warm-dense-matter, such as hydrodynamic, Monte-Carlo, or pseudo-potential approaches. 

It appears from figures \ref{fig1} to \ref{fig5} that none of the definitions we proposed for $Z^*$ in order to fulfill our requirements of continuity over large domains of densities and temperatures and coherence with the free electronic density of states assumed in Ziman's formalism, is able to achieve agreement with experiments or QMD calculations. Even with definition (\ref{def4}), a large discrepancy remains at $T\lesssim$ 15 000 K. More sophisticated average-atom models, based on more realistic ionic environments, give better results at low temperatures. Indeed, the SCAALP resistivities presented along figures \ref{fig1} to \ref{fig4}, however failing to present perfect quantitative agreement with experiments, give at least a relevant $\partial \eta/\partial T$ slope. Figure \ref{fig5} reports SCAALP results obtained for aluminum at $T=$ 10 000 K in the density range covering the metal to non-metal transition, and are represented by the green and black pluses. These two series of results differ by the way they account for the ion environment, through the ion-ion structure factor \cite{faussurier2006JQSRT,faussurier2006PRB}. Those represented by the black pluses were obtained with a sophisticated $S(q)$ resulting from the combination between SCAALP and Car-Parrinello Molecular-Dynamics, and are consistent with the bottom of the cloud of experimental values, whereas the green pluses, obtained with standard SCAALP model, are situated on the top of the cloud. These results illustrate the importance of solving self-consistently the electronic structure and the ion-ion structure factor at the average-atom code level. They suggest also the sensitivity to the model retained for $S(q)$ of beyond-jellium-approximation codes. Since all of these aspects are absent from our atom-in-jellium approach, we believe that agreement with experimental resistivities is most probably out of reach of any approach associating the Ziman-Evans formalism and standard atom-in-jellium average-atom method at low densities and low temperatures. In these conditions, definition (\ref{def4}) however offers the best possible agreement.\\

However, in Ref.~\cite{starrett2016}, Starrett pointed out the fact that the consistency of the Ziman formula would be improved if the density of states involved in the calculation of the resistivity was the same as the one used in the computation of the ionization. In the next section, we examine how the replacement, in the Ziman formulation, of the ideal electronic density of states by that given by {\sc Paradisio}, impacts the electrical resistivity, and if it is likely to modify significantly the main conclusions of the above comparisons with experimental data and other computations. We illustrate this discussion in the case of aluminum at 0.3 g.cm$^{-3}$ and 10 000 K, for which the average-atom continuum density of states departs the most from ideal one.

\section{Introduction of a non-ideal density of free states in Ziman's formalism}\label{sec4}

The expression of the Ziman-Evans resistivity can be rewritten as ($n_e$ being the electronic density, related to ionic one $n_i$ by $n_e=Z^* n_i$)
\begin{equation}
\eta=\dfrac{1}{n_e^2} \int_0^\infty d\epsilon \left(-\dfrac{\partial f}{\partial\epsilon}\right) \dfrac{(2\epsilon)^{3/2}}{3\pi^2} \dfrac{1}{\tau},
\end{equation}
introducing the inverse of the electron-ion relaxation time (we recall that $k^2=2\epsilon$, using the atomic units)
\begin{equation}
\dfrac{1}{\tau}=\pi n_i \dfrac{1}{k^3} \int_0^{2k} dq q^3 \dfrac{d\sigma}{d\theta} S(q)=\pi n_i \dfrac{\mathcal{I}(\epsilon)}{(2\epsilon)^{3/2}}.
\end{equation}
Noting that the quantity $(2\epsilon)^{3/2}/(3\pi^2)$ corresponds to the antiderivative of the UEG (ideal) density of state multiplied by the ion density $n_i$
\begin{equation}
\dfrac{(2\epsilon)^{3/2}}{3\pi^2}= n_i\,\int_0^\epsilon d\epsilon^\prime\,n^\mathrm{UEG}(\epsilon^\prime),
\end{equation}
with
\begin{equation}\label{UEGone}
n^\mathrm{UEG}(\epsilon)=\dfrac{(2\epsilon)^{1/2}}{\pi^2 n_i},
\end{equation}
the extension of Ziman's formula proposed in Refs.~\cite{starrett2016,potekhin1996} in order to account for any non-ideal density of states $n(\epsilon)$ reads 
\begin{align}
\eta&=\dfrac{1}{n_e^2} \int_0^\infty d\epsilon \left(-\dfrac{\partial f}{\partial\epsilon}\right) \mathcal{N}(\epsilon) \dfrac{1}{\tau}\nonumber\\&=\dfrac{\pi}{{Z^*}^2 n_i} \int_0^\infty d\epsilon \left(-\dfrac{\partial f}{\partial\epsilon}\right) \mathcal{N}(\epsilon) \dfrac{\mathcal{I}(\epsilon)}{(2\epsilon)^{3/2}}, 
\end{align}
where
\begin{equation}
\mathcal{N}(\epsilon)=n_i\,\int_0^\epsilon d\epsilon^\prime\,n(\epsilon^\prime).
\end{equation}
For the UEG density of states $n^\mathrm{UEG}(\epsilon)$ given by Eq.~(\ref{UEGone}), this function reads
\begin{equation}\label{UEGtwo}
\mathcal{N}^\mathrm{UEG}(\epsilon)=\dfrac{(2\epsilon)^{3/2}}{3 \pi^2}.
\end{equation}

\begin{figure}
\centering
\includegraphics[scale=0.40]{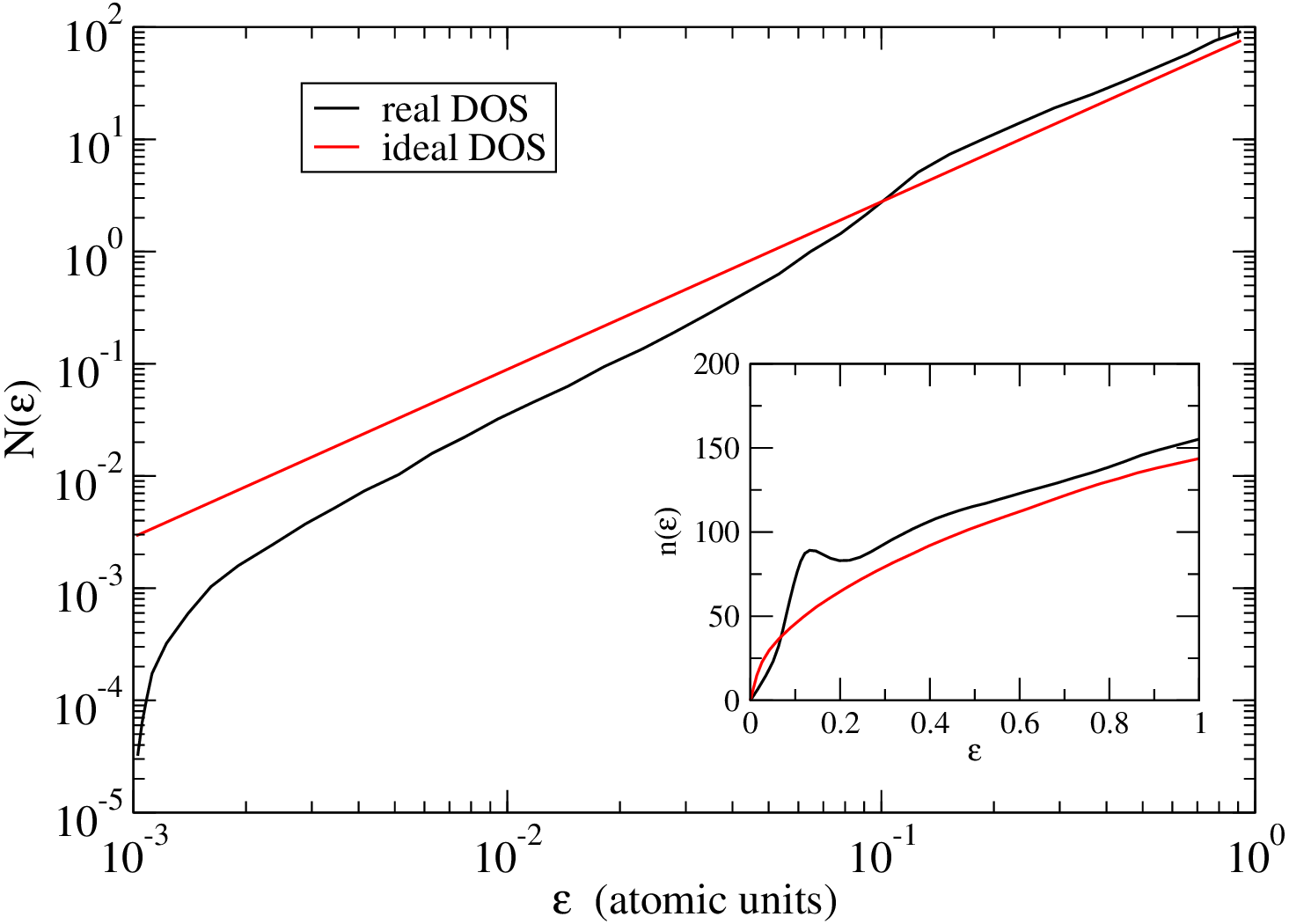}
\caption{Antiderivative of the density of free states (DOS): comparison between the ideal-gas case (given by Eq. (\ref{UEGtwo})) and the ``real'' one, obtained from a full-quantum mechanical average-atom calculation. The two latter density of states used to build $\mathcal{N}(\epsilon)$ are displayed in the insert.\label{fig11}}
\end{figure}

\begin{figure}
\centering
\vspace{1cm}
\includegraphics[scale=0.40]{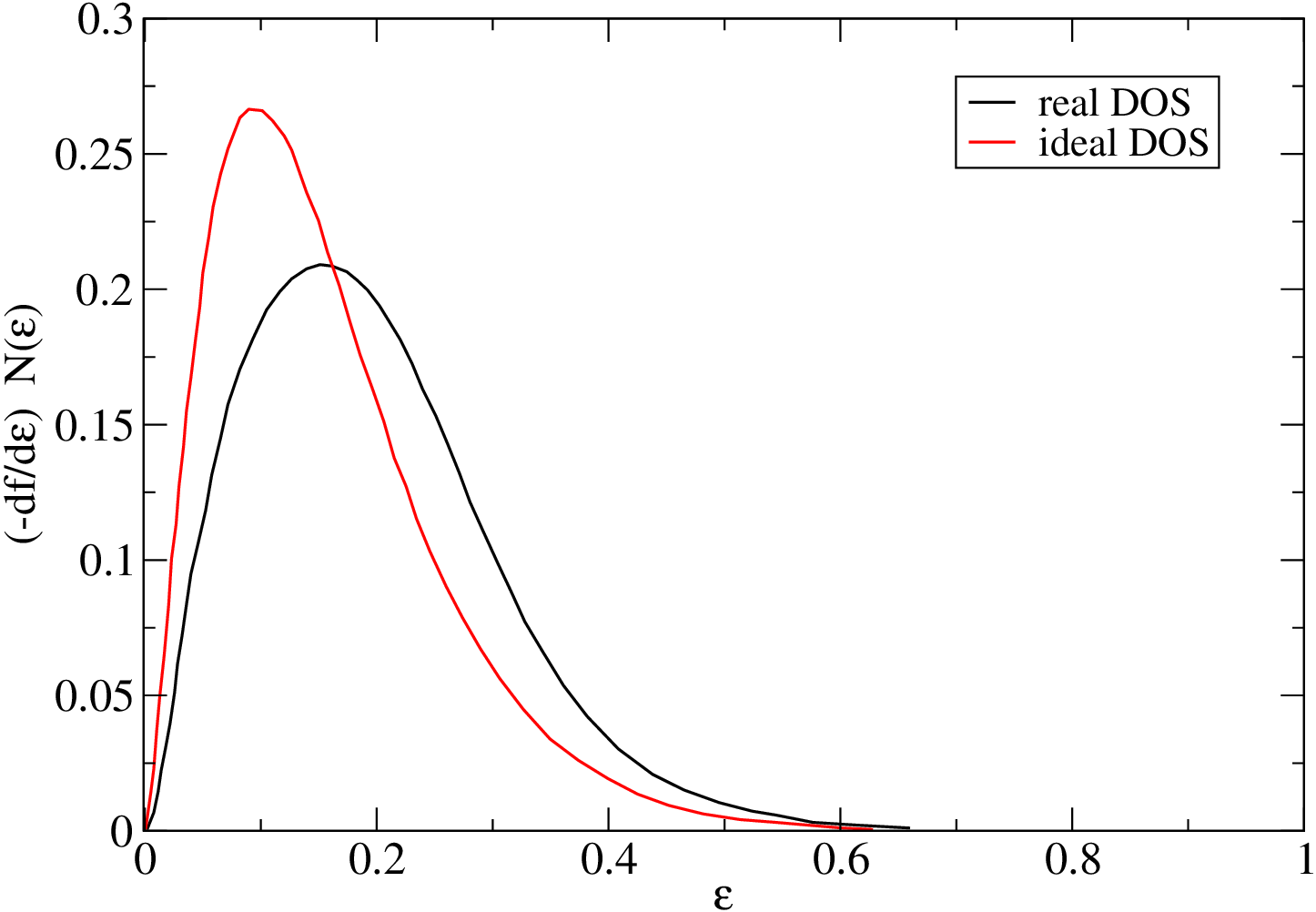}
\caption{Partial quantity $(-df/d\epsilon)~\mathcal{N}(\epsilon)$, involved in the Ziman formula and obtained from an average-atom computation, as a function of the continuum energy.\label{fig12}}
\end{figure}

Figure \ref{fig11} displays the functions $\mathcal{N}(\epsilon)$, {\it i.e.} $n_i$ times the antiderivatives of the real and ideal density of states $n(\epsilon)$ represented in the insert. In figure \ref{fig12} these two curves are multiplied by $(-\partial f/\partial\epsilon)$. At this stage, one expects a reduction of the resistivity when $Z^*=Z_\mathrm{cont}$ is used with the real continuum density of states.

\begin{figure}
\centering
\includegraphics[scale=0.40]{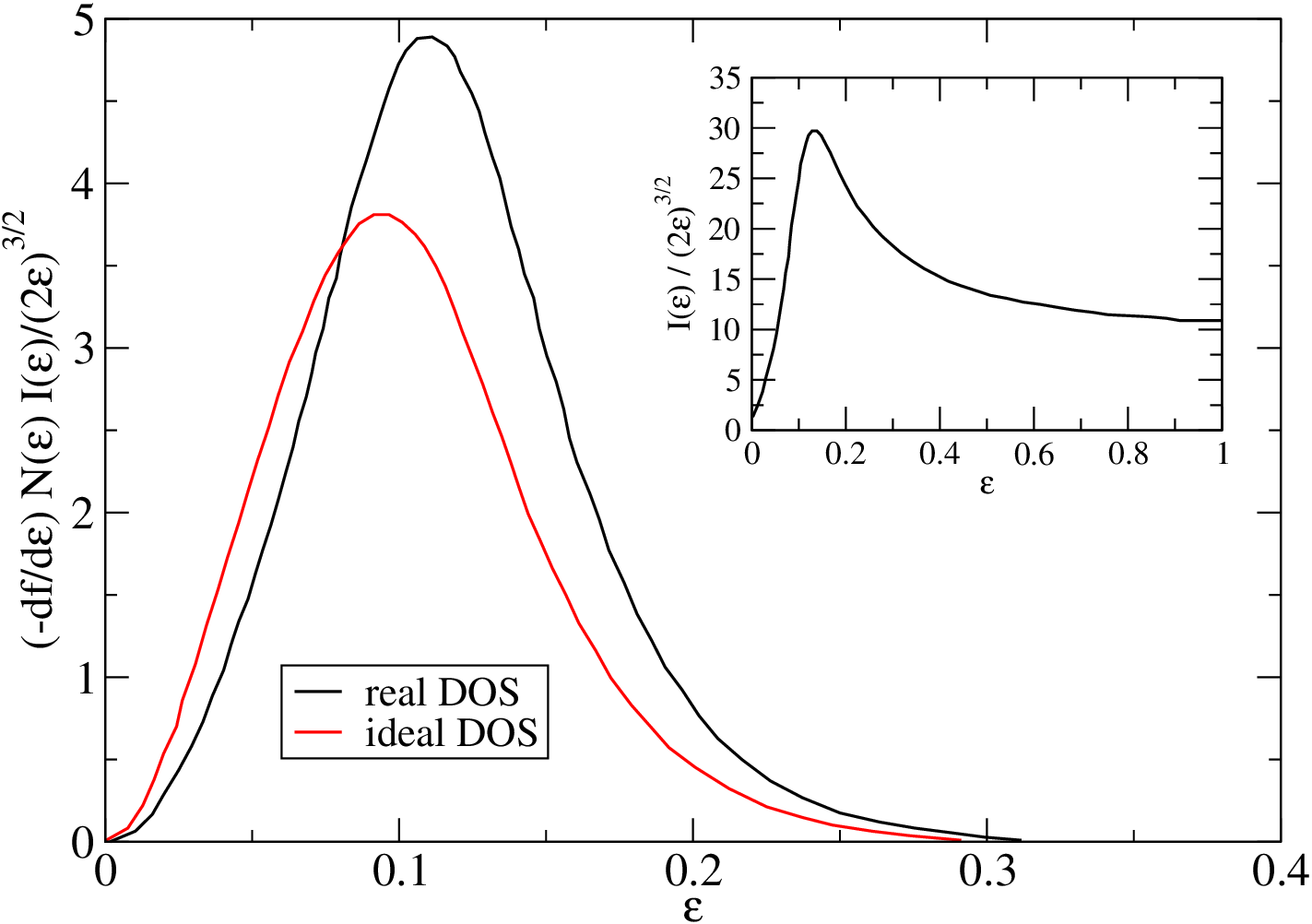}
\caption{Product of the two functions $(-df/d\epsilon)~\mathcal{N}(\epsilon)$ represented in figure \ref{fig12}, with the one $I(\epsilon)/(2\epsilon)^{3/2}$ shown in the insert, as a function of the continuum energy. The enhancement of this integrand when the real DOS is used in the modified Ziman formula increases the resistivity (and therefore reduces the conductivity).\label{fig14}}
\end{figure}

However, the quantities $\left(-\frac{\partial f}{\partial \epsilon} \right) \mathcal{N}(\epsilon)$ shown in figure \ref{fig12} must still be multiplied by the scattering contribution $\mathcal{I}(\epsilon)/(2\epsilon)^{3/2}$. This inverses the effect observed in figure \ref{fig12}. As shown in figure \ref{fig14}, the scattering contribution to the resistivity, represented in the insert, counters the effect of the introduction of the real density of states, which is consistent with an observation made by Ziman in his seminal review on the subject \cite{ziman1970}. In the present case of 0.3 g.cm$^{-3}$ aluminum at 10 000 K, the replacement of the ideal density of states by the real one induces an increase of the resistivity of the order of 25 \%. It is higher than the estimated 15 \% experimental error bars on the values of Ref.~\cite{clerouin2012}, but most experiments are more scattered than this latter uncertainty (see Figs.~\ref{fig4} and \ref{fig5}). The impact of the replacement of the ideal density of states by the real one is too limited to conclude that it will improve the agreement with experiments in the cases considered in this work. This is consistent with Ziman's conclusion in Ref. \cite{ziman1970}.

\section{Conclusion}\label{sec5}

We presented electrical resistivity calculations for boron, aluminum, titanium and copper expanded plasmas using the Ziman formulation in the framework of the average-atom model. We compared our results to experimental data, as well as other theoretical calculations, relying on the Ziman and Kubo-Greenwood formalisms, and based on average-atom models or quantum-molecular-dynamics simulations.

The ion charge $Z^*$ is a key parameter in Ziman's formalism, which identifies it to the charge of an ideal electronic conducting background. The difficulty is to determine which electronic states actually contribute to conductivity. Defining $Z^*$ as the number of perfectly unbound and free electrons, neglects less mobile ones, but however likely to contribute too. On the opposite, defining it as the number $Z_\mathrm{cont}$ of continuum electrons may include some charges that remain too close to the ion to contribute significantly to electrical conductivity, as it is the case at high densities, when resonances occur in the free density of states. For such cases, Petrov and Davidson proposed definition $Z^*=Z_\mathrm{cont}-Z_\mathrm{quasi-b}$, consisting in removing the number $Z_\mathrm{quasi-b}$ electrons trapped in the resonances from the total continuum electrons \cite{petrov2021}. In the present work, we proposed to replace $Z_\mathrm{quasi-b}$ by the charge $Z_F$ displaced by the ion-electron interaction, as given by Friedel's model of an impurity embedded in a perfect electron gas, from the total number $Z_\mathrm{cont}$ of conduction electrons. 

$Z_F$ gives the number of electrons that are the most strongly scattered by the central potential. For dense plasmas, it rises significantly in the conditions where resonances occur in the density of states, giving then the number of quasi-bound electrons trapped in these structures. Definition $Z^*=Z_\mathrm{cont}-Z_F$ tends then to the one proposed by Petrov and Davidson \cite{petrov2021}. However, ours offers more flexibility than Petrov and Davidson's one. Indeed, the displaced charge $Z_F$ can be positive or negative, depending whether the potential is attractive or repulsive. The latter case occurs for low-density plasmas. Some bound electrons become then ``quasi-free'' and contribute to the electrical conductivity. This adapts to the low-density case the concept of ``quasi-bound'' non conductive electrons originally applied to dense plasmas. Definition $Z^*=Z_\mathrm{cont}-Z_F$ generalizes then Petrov and Davidson's expression for the ion charge to low-density plasmas.

We found that such a description of electrons effectively contributing to the conductivity gives, within the Ziman theory combined with an average-atom computation of the atomic structure, the best possible agreement according to the jellium approximation for the environment, with measurements and quantum-molecular-dynamics simulations. At the opposite, the definitions $Z^*=Z_\mathrm{cont}$ and $Z^*=Z_\mathrm{free}+Z_F$ resulted in the less satisfying agreement. This could be due to an inconsistency between these latter two definitions and the assumption of an ideal density of free electronic states in Ziman's formula \cite{burrill2016,starrett2016}.

Therefore, we also considered the possibility of using $Z^*=Z_\mathrm{cont}$ in a modified Ziman formulation introducing explicitly a non-ideal density of states \cite{starrett2016,potekhin1996}. We showed that the positive impact on the resistivity induced in this way, is cancelled by the scattering contribution term in Ziman's formula, bringing back the final results within the experimental uncertainty.

Simultaneously with our study, Callow {\it et al.} were working on new ways to calculate mean ionization in the framework of average-atom models \cite{callow2023}. Their approach consists in partitioning bound and free electrons according to their shells, instead of their energies. One could here object that ours is based on the latter partitioning proposal, judged inappropriate by the authors. However, we can note that, by introducing the concepts of ``quasi-bound'' and ``quasi-free'' charges, our definition $Z^*=Z_\mathrm{cont}-Z_F$ alleviates the problem posed by differentiation based on energy sign.\\ 
In the present work, our aim was to propose a definition for $Z^*$ consistent with Ziman's approach for the electrical resistivity. Callow {\it et al.}'s approach is set in a more general framework, with the aim of proposing a global definition applicable to a large number of numerical methods involved in the study of the properties of warm dense matter, such as hydrodynamic or Monte-Carlo simulations, or pseudo-potential calculations. Although one of their proposals is closely related to the Kubo-Greenwood method, further investigations are necessary as concerns its interest for resistivity calculations within the Ziman formalism. This was out of the scope of the present work. We believe that the subject is interesting enough to deserve to be investigated separately. Their study was also limited to ambient densities.\\ 
Finally, as mentioned in their paper, their new methods for obtaining the mean ionization are not ``black-box'' ones, because of the need to set a boundary between bound and free orbitals. They therefore seem less easy to use for studies requiring calculations on a large range of thermodynamical conditions, than our proposal, which only requires the knowledge of the atomic number, the molar mass, a mass density and a temperature as input parameters and free from any adjustment.

\appendix
{
\section*{Appendix: the average-atom code Paradisio}
The {\sc Paradisio} \cite{penicaud2009} code is based on Liberman's relativistic quantum-average-atom model {\sc Inferno} \cite{liberman1979}. Within the latter, the atom is considered as a point nucleus surrounded by its $Z$ electrons, and placed at the center of a spherical cavity of radius $R_\mathrm{WS}$ embedded in a jellium. The Wigner-Seitz radius $R_\mathrm{WS}$ reads
\begin{equation}\label{Rws}
R_\mathrm{WS}=\left(\dfrac{3}{4\pi}\dfrac{A/\mathcal{N}_\mathrm{Avo}}{\rho} \right)^{1/3},
\end{equation}
$\rho$, $A$ and $\mathcal{N}_\mathrm{Avo}$ denoting respectively the mass density, molar mass and Avogrado number.

The jellium takes place of the surrounding ions, and consists in a uniform electron gas and a uniform distribution of positive charges that ensures its electrical neutrality. Electrical neutrality inside the cavity is imposed too. {\sc Paradisio} computes then electronic structure in a self-consistent way. The only required parameters are atomic number $Z$, molar mass $A$, mass density $\rho$ and temperature $T$.

Atomic units where $e=\hbar=m=1$, and where the celerity of light $c=137.036$ is the inverse of the fine structure constant $\alpha=e^2/(8\pi\epsilon_0 a_B)$, $a_B$ being the Bohr radius and $\epsilon_0$ the permittivity of vacuum, are used throughout the appendix.

In a spherically symmetric potential, the one-electron wave-functions, solutions of Dirac equation, are of the form
\begin{equation}\label{psis}
\psi_s(\vec{r})\equiv\psi_{j\ell m}(\vec{r})=\left(\begin{array}{l}
\frac{1}{r}F(r)\Omega_{j\ell m}(\theta,\phi)\\
-\frac{i}{r}G(r)\Omega_{j\ell'm}(\theta,\phi)
\end{array}
\right),
\end{equation}
where $\Omega_{j\ell m}$ and $\Omega_{j'\ell m}$ are two spinors. $j$, $\ell$ and $m$ are quantum numbers associated respectively to the total angular momentum $J$, to the orbital angular momentum $L$ and its projection $L_z$ on the $z$ axis. The quantum number $\ell'$ is given by: 
\begin{equation}
\ell'=\left\{\begin{array}{ll}
\ell+1\;\;\;\;\mathrm{if}\;\;\;\; j=\ell+1/2\\
\ell-1\;\;\;\;\mathrm{if}\;\;\;\; j=\ell-1/2.
\end{array}
\right.
\end{equation}

The Dirac equation reduces then to the following equations verified by the radial functions $F(r)$ and $G(r)$ 
\begin{equation}\label{radial}
\left\{
\begin{array}{l}
\dfrac{dF}{dr}=-\dfrac{\kappa}{r}F(r)-\dfrac{V_{\mathrm{eff}}(r)-c^2-\epsilon}{c}G(r)\\
\\
\dfrac{dG}{dr}=\dfrac{V_{\mathrm{eff}}(r)+c^2-\epsilon}{c}F(r)+\dfrac{\kappa}{r}G(r)
\end{array}
\right.
\end{equation}
where
\begin{equation}
\left\{
\begin{array}{l}
\kappa=-(\ell+1)\;\;\;\;\mathrm{for}\;\;\;\; j=\ell+1/2,\\
\kappa=\ell\;\;\;\;\;\;\;\;\;\;\;\;\;\;\;\;\mathrm{for}\;\;\;\; j=\ell-1/2.
\end{array}
\right.
\end{equation}

The effective potential $V_\mathrm{eff}(r)$ reads
\begin{eqnarray}\label{pottot}
\left\{
\begin{array}{l}
V_{\mathrm{eff}}(r)=V_c(r)+V_{\mathrm{xc}}(r)-\nu\;\;\;\;\mathrm{if}\;\;\;\;r\leq R_\mathrm{WS},\\
V_{\mathrm{eff}}(r)=V_{\infty}\;\;\;\;\;\;\;\;\;\;\;\;\;\;\;\;\;\;\;\;\;\;\;\;\;\;\;\mathrm{if}\;\;\;\;r>R_\mathrm{WS},
\end{array}
\right.
\end{eqnarray} 
where $V_c(r)$ is the Coulomb potential
\begin{equation}
V_c(r)=-\frac{Z}{r}+\int_{r'\leq R_\mathrm{WS}}\frac{n(r')}{|\vec{r}-\vec{r'}|}\, 4\pi {r'}^2 dr',
\end{equation}
$n(r)$ denoting the electronic density. $V_\mathrm{xc}(r)$ is the exchange-correlation potential, equal to the exchange-correlation chemical potential evaluated at local density $n(r)$: 
\begin{equation}
V_{\mathrm{xc}}(r)=\mu_\mathrm{xc}[n(r),T],
\end{equation}
the exchange-correlation chemical potential $\mu_\mathrm{xc}[n,T]$ being related to the exchange-correlation free energy functional $f_\mathrm{xc}$ functional by
\begin{equation}
\mu_\mathrm{xc}[n,T]=\left.\frac{\partial}{\partial n}nf_{\mathrm{xc}}[n,T]\right|_T.
\end{equation}
{\sc Paradisio} uses Karasiev {\it et al}'s Pad\'e approximants \cite{karasiev2014}, with Groth {\it et al.}'s revised parameters \cite{groth2017}. The quantity $\nu$ in Eq.~(\ref{pottot}) is
\begin{align}
\nu=&f_{\mathrm{xc}}[n(R_\mathrm{WS}),T]-\mu_{\mathrm{xc}}[\bar{n},T]\\
&+\frac{\bar{n}}{n(R_\mathrm{WS})}\left(\mu_{\mathrm{xc}}[\bar{n},T]-f_{\mathrm{xc}}[\bar{n},T]\right),\nonumber
\end{align}
where $n(R_\mathrm{WS})$ is the electron density at the radius of the cavity. Finally, the value of the potential outside cavity, denoted $V_\infty$, is 
\begin{equation}
V_{\infty}=\mu_{\mathrm{xc}}[\bar{n},T],
\end{equation}
$\bar{n}$ denoting the density of the jellium. 

The solutions of Eqs.~(\ref{radial}) for $r>R_\mathrm{WS}$ where the potential is constant are known, and therefore, Eqs.~(\ref{radial}) only have to be solved for $r\leq R_\mathrm{WS}$. The inside and outside solutions are matched at $r=R_\mathrm{WS}$.

The model imposes $F(r)=G(r)=0$ at $r=0$ and $r\rightarrow\infty$. Outside the cavity, the radial functions $F^\mathrm{oc}(r)$ and $G^\mathrm{oc}(r)$ (the superscript ``oc'' stands for ``outside cavity'') satisfying those boundary conditions are, for bound states, modified Bessel functions of the third kind \cite{abramowitz1965}, exponentially decreasing, and, for free states, combinations of Bessel functions of the first and second kinds, with decreasing amplitudes as $r\rightarrow\infty$.

The solutions of Eqs.~(\ref{radial}) are, for $r\geq R_\mathrm{WS}$:
\begin{itemize}
\item For $\epsilon<V_{\infty}$ (bound states):
\end{itemize}
\begin{equation}
\left\{
\begin{array}{l}
F^\mathrm{oc}_\mathrm{b}(r)=a_0c\frac{k}{V_{\infty}-\epsilon}rK_{\ell+1/2}(kr)\\
\\
G^\mathrm{oc}_\mathrm{b}(r)=a_0rK_{\ell'+1/2}(kr),
\end{array}
\right.
\end{equation}
where the $K_{n+1/2}$ ($n$ being an integer) are modified Bessel functions of the third kind and $a_0$ the normalization constant
\begin{equation}
a_0=\frac{1}{\displaystyle\int_0^{\infty}\left\{\left[F^\mathrm{oc}_\mathrm{b}(r)\right]^2+\left[G^\mathrm{oc}_\mathrm{b}(r)\right]^2\right\} dr}.
\end{equation}
These functions connect to inside cavity ones only for some values of $\epsilon$, yielding the discrete set of bound energies.

\begin{itemize}
\item For $\epsilon>V_{\infty}$ (free states):
\end{itemize}
\begin{equation}
\left\{
\begin{array}{l}
F^\mathrm{oc}_\mathrm{f}(r)=b_0c\frac{k}{\epsilon-V_{\infty}}r\left[\cos(\delta_\kappa) j_{\ell}(kr)-\sin(\delta_\kappa) n_{\ell}(kr)\right]\\
G^\mathrm{oc}_\mathrm{f}(r)=b_0r\left[\cos(\delta_\kappa) j_{\ell'}(kr)-\sin(\delta_\kappa) n_{\ell'}(kr)\right],
\end{array}
\right.
\end{equation}
where the normalization factor $b_0$ reads 
\begin{equation}
 b_0=\sqrt{\frac{2}{\pi}}\displaystyle\frac{k}{\sqrt{1+\displaystyle\frac{c^2k^2}{\left(\epsilon-V_{\infty}\right)^2}}}.
\end{equation}
The matching of these outside cavity radial functions with the internal ones at $r=R_\mathrm{WS}$ is always possible by adjusting the phase-shifts $\delta_\kappa(k)$. 

The wave number $k$ and the energy $\epsilon$ are related by:
\begin{equation}
k=\sqrt{2\left(V_{\infty}-\epsilon\right)\left[1-\frac{\left(V_{\infty}-\epsilon\right)}{2c^2}\right]}.
\end{equation}

The electronic density $n(r)$ follows then
\begin{align}
n(r)=&\sum_\mathrm{b} \sum_\kappa 2|\kappa|\,\left[F_\mathrm{b}(r,\kappa,\epsilon_\mathrm{b})^2+G_\mathrm{b}(r,\kappa,\epsilon_\mathrm{b})^2 \right]\nonumber\\
&+\int_0^\infty d\epsilon\, \sum_\kappa 2|\kappa|\,\left[F_\mathrm{f}(r,\kappa,\epsilon)^2+G_\mathrm{f}(r,\kappa,\epsilon)^2 \right].
\end{align}

The number $Z_\mathrm{bound}$ of bound electrons and the number $Z_\mathrm{cont}$ of continuum ones respectively read
\begin{equation}
Z_{\mathrm{bound}}=\sum_ \mathrm{b} f(\epsilon_\mathrm{b},\mu) \sum_\kappa 2|\kappa|\,\left\{\int_0^{R_\mathrm{WS}}\left[F_\mathrm{b}(r,\kappa,\epsilon_\mathrm{b})^2+G_\mathrm{b}(r,\kappa,\epsilon_\mathrm{b})^2 \right]\,r^2 dr \right\},
\end{equation}
\begin{equation}
Z_{\mathrm{cont}}=\int_0^\infty d\epsilon f(\epsilon,\mu) \sum_\kappa 2|\kappa|\,\left\{\int_0^{R_\mathrm{WS}}\left[F_\mathrm{f}(r,\kappa,\epsilon)^2+G_\mathrm{f}(r,\kappa,\epsilon)^2 \right]\,r^2 dr \right\}.
\end{equation}
$f(\epsilon,\mu)$ is the Fermi-Dirac energy distribution
\begin{equation}
f(\epsilon,\mu)=\dfrac{1}{\mathrm{e}^{\beta(\epsilon-\mu)}+1},
\end{equation}
where $\beta=1/T$, and the chemical potential $\mu$ is determined from the electro-neutrality condition
\begin{equation}
Z_{\mathrm{bound}}+Z_{\mathrm{cont}}=Z.
\end{equation}
}
Finally, the jellium density $\overline{n}$ is 
\begin{equation}
\overline{n}=\dfrac{\sqrt{2}}{\pi^2 \beta^{3/2}}\mathcal{F}_{1/2}(\beta\mu),
\end{equation}
where 
\begin{equation}
\mathcal{F}_{1/2}(x)=\int_0^\infty dt \dfrac{t^{1/2}}{e^{t-x}+1}
\end{equation}
defines the Fermi function of order 1/2.


\end{document}